\documentclass[superscriptaddress, nofootinbib, preprintnumbers, twocolumn, amsmath,amssymb,aps]{revtex4-2}

\usepackage{amsmath}	% AMS Math Package
\usepackage{amsthm}		% Theorem Formatting
\usepackage{amssymb}	% Math symbols such as \mathbb
\usepackage{eufrak}
\usepackage{datetime}
\usepackage{graphicx}
\usepackage{color}
\usepackage{verbatim}
\usepackage{bm}

\usepackage[colorlinks=true, citecolor=midblue, linkcolor=midblue, urlcolor=midblue]{hyperref}

\usepackage{setspace}
\usepackage[flushleft]{threeparttable}
\usepackage{multirow}
\usepackage{tabularx, booktabs}
\usepackage{array}
\newcolumntype{?}{!{\vrule width 1pt}}

\definecolor{grey}{rgb}{0.4,0.4,0.4}
\definecolor{dullmagenta}{rgb}{0.4,0,0.4}
\definecolor{darkblue}{rgb}{0,0,0.4}
\definecolor{midblue}{rgb}{0,0,0.5}
\definecolor{midred}{rgb}{0.5,0,0}
\definecolor{orange}{rgb}{1,0.5,0}
\definecolor{lightbrown}{rgb}{0.75,0.5,0.25}
\definecolor{tan}{cmyk}{0.14,0.42,0.56,0}
\definecolor{djunglegreen}{cmyk}{0.99,0,0.52,0}
\definecolor{lightgreen}{rgb}{0,1,0}
\definecolor{olivegreen}{cmyk}{0.64,0,0.95,0.40}
\definecolor{midgreen}{rgb}{0.0,0.675,0.0}
\definecolor{darkgreen}{rgb}{0,0.5,0}

\newcommand{\vs}{\vspace}

\newcommand{\Urm}{\ensuremath{\mathrm{U}}}

\newcommand{\crm}{\ensuremath{\mathrm{c}}}

\newcommand{\erm}{\ensuremath{\mathrm{e}}}

\newcommand{\hrm}{\ensuremath{\mathrm{h}}}

\newcommand{\QCD}{{\rm QCD}}
\newcommand{\PQ}{{\rm PQ}}

\newcommand{\DW}{{\rm DW}}

\newcommand{\GeV}{\ensuremath{\, \mathrm{GeV}}}

\smallskipamount
\settimeformat{ampmtime}

\usepackage{float}

\newcolumntype{Y}{>{\centering\arraybackslash}X}
\newcolumntype{Z}{>{\raggedleft\arraybackslash}X}

\newcommand{\Haloscopes}{\cite{Beurthey:2020yuq, Grenet:2021vbb, CAST:2020rlf, Crisosto:2019fcj, Devlin:2021fpq, QUAX:2020adt, Alesini:2019ajt, Alesini:2020vny, Thomson:2019aht, McAllister:2017lkb, Jeong:2020cwz, Gramolin:2020ict, Salemi:2021gck, Ouellet:2018beu, HAYSTAC:2018rwy, HAYSTAC:2020kwv, hagmann1990results, depanfilis1987limits, ADMX:2018gho, ADMX:2019uok, ADMX:2021nhd, Baryakhtar:2018doz, Michimura:2019qxr, Aja:2022csb, BREAD:2021tpx, Millar:2022peq, Lawson:2019brd, TASEH:2022vvu, 2022PhRvD.106e2007A, 2022arXiv220312152Q, 2023arXiv230109721H, 2020PhRvL.124j1802L, 2021PhRvL.126s1802K, 2022PhRvL.128x1805L, 2022arXiv220713597K, 2022arXiv221010961Y, Berlin:2020vrk, DMRadio:2022pkf, Alesini:2017ifp, Zhang:2021bpa, Nagano:2019rbw, Liu:2018icu, Schutte-Engel:2021bqm, Marsh:2018dlj}}

\makeatletter

\def\@sect@ltx#1#2#3#4#5#6[#7]#8{%
    \@ifnum{#2>\c@secnumdepth}{%
        \def\H@svsec{\phantomsection}%
        \let\@svsec\@empty
    }{%
        \H@refstepcounter{#1}%
        \def\H@svsec{%
            \phantomsection
        }%
        \protected@edef\@svsec{{#1}}%
        \@ifundefined{@#1cntformat}{%
            \prepdef\@svsec\@seccntformat
        }{%
            \expandafter\prepdef
            \expandafter\@svsec
            \csname @#1cntformat\endcsname
        }%
    }%
    \@tempskipa #5\relax
    \@ifdim{\@tempskipa>\z@}{%
        \begingroup
        \interlinepenalty \@M
        #6{%
            \@ifundefined{@hangfrom@#1}{\@hang@from}{\csname @hangfrom@#1\endcsname}%
            {\hskip#3\relax\H@svsec}{\@svsec}{#8}%
        }%
        \@@par
        \endgroup
        \@ifundefined{#1mark}{\@gobble}{\csname #1mark\endcsname}{#7}%
        \addcontentsline{toc}{#1}{%
            \@ifnum{#2>\c@secnumdepth}{%
                \protect\numberline{}%
            }{%
                \protect\numberline{\csname the#1\endcsname}%
            }%
            #7}% <<<<<<<<<<<<<<<<<<<<<<<< changed from #8
    }{%
        \def\@svsechd{%
            #6{%
                \@ifundefined{@runin@to@#1}{\@runin@to}{\csname @runin@to@#1\endcsname}%
                {\hskip#3\relax\H@svsec}{\@svsec}{#8}%
            }%
            \@ifundefined{#1mark}{\@gobble}{\csname #1mark\endcsname}{#7}%
            \addcontentsline{toc}{#1}{%
                \@ifnum{#2>\c@secnumdepth}{%
                    \protect\numberline{}%
                }{%
                    \protect\numberline{\csname the#1\endcsname}%
                }%
                #8}%
        }%
    }%
    \@xsect{#5}}%

\makeatother
\begin{document}

%%%%%%%%%%%%%%%%%%%%%%%%%%%%%%%%%%%%%%%%%%%%%%%%%%%%%%%%
\title{DFSZ-Type Axions and Where to Find Them}

\author{Johannes Diehl}
\email{diehl@mpp.mpg.de}
\affiliation{
	Max-Planck-Institut f{\"u}r Physik,
	F{\"o}hringer Ring 6,
	80805 M{\"u}nchen,
	Germany}
	
\author{Emmanouil Koutsangelas}
\email{emi@mpp.mpg.de}
\affiliation{
	Arnold Sommerfeld Center,
	Ludwig-Maximilians-Universit{\"a}t M{\"u}nchen,
	Theresienstra{\ss}e 37,
	80333 M{\"u}nchen,
	Germany}
\affiliation{
	Max-Planck-Institut f{\"u}r Physik,
	F{\"o}hringer Ring 6,
	80805 M{\"u}nchen,
	Germany}

\date{\formatdate{\day}{\month}{\year}, \currenttime}

%%%%%%%%%%%%%%%%%%%%%%%%%%%%%%%%%%%%%%%%%%%%%%%%%%%%%%%%
\begin{abstract}

We systematically calculate the axion-photon coupling for non-minimal DFSZ models. Thereby we can classify every calculated model and study the resulting distributions, relevant for axion experiments like haloscopes, helioscopes or light-shining-through-a-wall experiments. By adding more than one additional Higgs doublet, these non-minimal DFSZ models extend the viable axion parameter space and lead to a large range of axion-photon couplings. We find couplings almost three orders of magnitude larger than the ones of the minimal models. Most of the possible axion-photon couplings, however, lie in the vicinity of the values dictated by the minimal models. We quantify this by introducing a theoretical prior probability distribution for DFSZ-type axions and giving $68\%$ and $95\%$ lower bounds as well as two-sided bands. 
We compare our results for the DFSZ axion-photon coupling distributions with the KSVZ case, for which a similar analysis has been conducted. Both display similar values as well as a very specific pattern.
In order to identify preferred models, we discuss the role of flavour changing neutral currents and the domain wall problem as possible selection criteria. It is possible to construct a large number of non-minimal DFSZ models with a domain wall number of unity, thereby avoiding the domain wall problem. This subset also has a significantly enhanced axion-photon coupling compared to the minimal DFSZ models.

\end{abstract}

%%%%%%%%%%%%%%%%%%%%%%%%%%%%%%%%%%%%%%%%%%%%%%%%%%%%%%%%
\maketitle

\tableofcontents

%%%%%%%%%%%%%%%%%%%%%%%%%%%%%%%%%%%%%%%%%%%%%%%%%%%%%%%%
%%%%%%%%%%%%%%%%%%%%%%%%%%%%%%%%%%%%%%%%%%%%%%%%%%%%%%%%
\section[INTRODUCTION]{INTRODUCTION}
\label{sec:Introduction}
\vs{-5mm}

The strong CP problem remains one of the biggest puzzles of particle physics. While it is usually expressed as the inexplicable smallness of the CP violating $\theta$-parameter, i.e.~$\theta < 10^{-10}$ \cite{nEDM2015}, it is in fact a vacuum selection problem rooted in the non-trivial vacuum structure of QCD \cite{Weinberg:1996kr}. What makes this ``small value" problem special is that within the Standard Model (SM) quantum corrections to $\theta$ are many orders of magnitude below the experimental bound \cite{ThetaCorrections} (unlike the Higgs mass for instance) so that it is not really a problem for $\theta$ to be that small. In this sense any explanation for the smallness of $\theta$ is just theoretically motivated.

However, in recent years it has been pointed out that when quantum gravity is taken into account, consistency relations are imposed that exclude any type of (meta) stable de Sitter vacua  \cite{Dvali:2013eja, Dvali:2018jhn, Dvali:2020etd}. Since any theory with $\theta \neq 0$ is of de Sitter-type, this reduces the number of viable vacua to exactly one: the CP conserving vacuum at $\theta=0$. This does not only promote the strong CP problem to a real problem but it also makes a mechanism that results in $\theta=0$ a necessity \cite{Dvali:2018dce, Dvali:2022fdv}.

Such a mechanism is given by the Peccei-Quinn (PQ) solution, which essentially introduces a non-linearly realised $\Urm(1)_\PQ$ that is anomalous with respect to QCD \cite{PQMechanism, PQMechanism2}. The crucial anomaly condition can be expressed as the non-conservation of the PQ current, namely
    \begin{equation}
        \partial_\mu J_\PQ^\mu
            =
                N \frac{\alpha_s}{4\pi} G \tilde{G}
                + 
                E \frac{\alpha}{4\pi} F \tilde{F}
            \; ,
    \label{Eq:PQAnomaly}
    \end{equation}
where the electromagnetic anomaly, which in general is also present, is added. In this expression $G, F$ denote the colour and electromagnetic field strength tensors, $\tilde{G}, \tilde{F} $ their duals, $\alpha_s , \alpha$ their associated fine-structure constants, and $N, E$ the corresponding anomaly coefficients. The PQ mechanism solves the strong CP problem by making the $\theta$-parameter unobservable as it gets relaxed to zero by the pseudo-Goldstone boson of the PQ symmetry, the axion \cite{WeinbergAxion, Wilczek:1977pj}.

The PQ solution is special in the sense that it predicts a new light pseudoscalar particle. However, it does not specify the axion low-energy couplings, which depend on the UV physics \cite{SREDNICKI1985689}. The low-energy effective theory is thus not sufficient, and UV models are needed to make concrete predictions about the couplings of the axion. This is usually achieved by the two large classes of invisible axion models, the DFSZ-type \cite{DFSZ1, DFSZ2} and KSVZ-type models \cite{KSVZ1, KSVZ2}. The former adds Higgs singlets and doublets to the SM, while the latter adds Higgs singlets and heavy quarks (for a review see \cite{DiLuzioAxionLandscape}). Even though the minimal models of each type, adding only one Higgs doublet or one quark of arbitrary representation, define benchmark models, in principle there is a plethora of non-minimal models. An identification of all these models and systematic approach that allows to extract a prediction from all these models at the same time would be desirable. The goal of this work is to do exactly this for the DFSZ-type axions.

We achieve this by exploiting the unique property of the axion-photon coupling: Its UV physics are fully encoded in the ratio between the electromagnetic and the QCD anomaly coefficients \cite{SREDNICKI1985689},
    \begin{equation}
        g_{a\gamma}
            =
                \frac{\alpha}{2\pi f_a}
                \left[
                    \frac{E}{N}
                    - 1.92(4)
                \right]
            \equiv
                \frac{\alpha}{2\pi f_a}
                \mathcal{C}_{a\gamma}
            \; ,
    \label{Eq:AxionPhotonC}
    \end{equation}
where $f_a$ is the axion decay constant and by $\mathcal{C}_{a\gamma}$ we denote the dimensionless part.

Because of the nature of anomalies, this ratio does not depend on unknown vacuum expectation values (VEVs) or mixing angles but only on the representation of the fields. For the DFSZ-type models, this comes down to fixing the PQ charges of the SM fermions, which are not free but determined by consistency and phenomenology conditions \cite{DiLuzioAxionLandscape}. Systematically solving the associated linear system of equations (LSE) allows us to calculate the anomaly ratio and thus the axion-photon coupling for a large number of DFSZ-type models.

In addition to calculating the anomaly ratios for a large number of DFSZ-type models, we are able to count how many different models lead to the same anomaly ratio. We then use this notion of multiplicity to allocate a certain probability to each anomaly ratio. From analysing the resulting distributions we are able to extract several interesting conclusions for the axion experimental program. One of our key observations is that the values dictated by the minimal DFSZ models, namely $E/N = 2/3$ and $E/N=8/3$, are statistically favoured for DFSZ-type theories, even though for a larger number of Higgs doublets many more possible values for the anomaly ratio can exist. While this confirms the potential experimental importance of these values, on the other hand we argue that a non-observation at these values still leaves a significant amount of the axion parameter space viable. We quantify this statement by defining an axion band as well as lower $g_{a\gamma}$ bounds.

A similar analysis has already been done for the KSVZ-type axion: The identification and classification is described in \cite{DiLuzio:2017pfr}, while the statistical analysis is reported in \cite{Plakkot}. Furthermore, in \cite{DiLuzio:2017pfr} a part is dedicated to DFSZ-type axions. There, by estimating the maximal possible anomaly ratio, the authors argue that the majority of realistic DFSZ-type models lie in the same window as the preferred KSVZ-type ones. With our work, we are not only able to give a more precise value of the maximal possible anomaly ratio, which turns out to be higher than the previous estimate, but to also perform a detailed comparison between the two classes, which allows us to better understand their relation.

The paper is organised as follows. To begin with, in Sec.~\ref{sec:DFSZ-TYPE_OF_AXION_MODEL} we review DFSZ-type models where we put the focus on the determination of the PQ charges. Moreover, we discuss potential phenomenological selection criteria and give a general procedure on how to determine all possible anomaly ratios and their multiplicities for a given number of Higgs doublets. In Sec.~\ref{sec:ANOMALY_RATIO_DISTRIBUTIONS} we apply this approach to theories with three to nine Higgs doublets. We discuss arising problems for a high number of doublets and compare our results with the KSVZ-type models. 
Next, Sec.~\ref{sec:IMPACT_ON_AXION_SEARCHES} discusses experimental implications by estimating necessary sensitivities for axion searches. Lastly, in Sec.~\ref{sec:Summary_and_Outlook} we summarise our results and give an outlook.

%%%%%%%%%%%%%%%%%%%%%%%%%%%%%%%%%%%%%%%%%%%%%%%%%%%%%%%%
%%%%%%%%%%%%%%%%%%%%%%%%%%%%%%%%%%%%%%%%%%%%%%%%%%%%%%%%
\section{DFSZ-TYPE AXION MODELS}
\label{sec:DFSZ-TYPE_OF_AXION_MODEL}
\vs{-5mm}

In the DFSZ-type of models, the fermionic fields of the SM are charged under the PQ symmetry. This requires to enlarge the scalar content of the SM by one singlet and at least one additional Higgs doublet. The additional doublets are required for the PQ mechanism to make the PQ symmetry anomalous with respect to QCD. The singlet is introduced to render the axion invisible by decoupling the PQ scale from the electroweak scale \cite{DFSZ1,DFSZ2}. 

The anomaly of the PQ current only depends on the difference between the PQ charges of left- and right-handed fermions. For simplicity we set the PQ charges of the left-handed fermions to zero. This leaves us with the charges of the right-handed ones, which we denote as $\chi_{u_i}, \chi_{d_i}, \chi_{e_i}$ with $i$ being a generation index. Regarding the neutrinos the situation is somewhat special. While the left-handed neutrino is not directly contributing to the anomaly ratio $E/N$, it could be contributing indirectly if a right-handed neutrino was present in the theory. We mean by this that by having a right-handed neutrino, we can write a Dirac mass term for the neutrinos with another Higgs doublet. This Higgs doublet can then be used to change the PQ charges of the other fermions and thus, indirectly, the anomaly ratio. Since it is currently unknown if the neutrino masses are realised via the type-I seesaw mechanism, which requires the introduction of the right-handed neutrinos, we exclude the neutrinos in our analysis by setting their PQ charge to zero in accordance with the other left-handed fermions.

%What we mean is that by having a right-handed neutrino, we can write a Dirac mass term for the neutrinos with another Higgs doublet. The neutrinos do not contribute into the anomaly ratio, however, with another Higgs doublet we can write explicit breaking terms in the potential that change the PQ charges of the other fermions. Di Luzio mentions this in his paper 1705.05370 in section VIII  with doublets that do not couple to fermions. With a right-handed neutrino, we can have a similar effect with a coupling to fermions.

In the following we denote the DFSZ-type models as DFSZ$_{n_D}$, where $n_D$ is the total number of doublets. In this terminology the original models, which represent the minimal versions, become DFSZ$_2$-I and DFSZ$_2$-II with $E/N = 2/3$ and $E/N = 8/3$, respectively. 

%%%%%%%%%%%%%%%%%%%%%%%%%%%%%%%%%%%%%%%%%%%%%%%%%%%%%%%%
%%%%%%%%%%%%%%%%%%%%%%%%%%%%%%%%%%%%%%%%%%%%%%%%%%%%%%%%
\subsection{Identifying the Axion}
\label{Subsec:Identifying_the_Axion}
\vs{-5mm}

Let us for concreteness consider a DFSZ$_{n_D}$ model with $n_D \leq 9$ and begin by fixing the Yukawa sector. In order to exhaust the maximum freedom of the PQ charges, we consider a Yukawa sector where each right-handed fermion couples to only one doublet. This makes it reasonable to denote the doublets as $H^{u_i}, H^{d_i}, H^{e_i},$ and the singlet as $S$. The Yukawa sector then takes the form
	\begin{equation}
		\mathcal{L}
			\supset
				- y^u_{ij} H^{u_i} \bar{Q}_L^i u_R^j
				- y^d_{ij} H^{d_i} \bar{Q}_L^i d_R^j
				- y^e_{ij} H^{e_i} \bar{E}_L^i e_R^j
				+ \hrm.\crm.
			\; .
		\label{Eq:Yukawa}
	\end{equation}
For $n_D = 9$ each right-handed fermion couples to a different doublet, while for $n_D < 9$ some fermions have to couple to the same doublet. This form of the Yukawa sector automatically fixes the weak hypercharge of the doublets to be
    \begin{equation}
        - Y_{H^{u_i}} = Y_{H^{d_i}} = Y_{H^{e_i}} = \frac{1}{2} \; .
    \end{equation}
In principle, several doublets can couple to the same right-handed fermion. We ignore this issue for now and come back to it in Sec.~\ref{Subsec:Multiplicity}.

Next, the standard kinetic term for each scalar is invariant under a $\Urm(1)^{n_D+1}$ symmetry. This symmetry must be explicitly broken down to $\Urm(1)_{\rm PQ} \times \Urm(1)_Y$ for the PQ current to be well-defined and to avoid Goldstone bosons with decay constants of electroweak scale order. With this requirement in mind, we split the potential into two parts,
    \begin{equation}
        V 
            = 
                V_{\rm moduli}
                + V_{\rm eb}
            \; .
    \end{equation}
The first term, $V_{\rm moduli}$, only consists of the modulus of each scalar or the modulus of two doublets and hence does not break any of the global U(1) groups explicitly. In contrast, $V_{\rm eb}$ consists of terms that all break the $\Urm(1)^{n_D+1}$ symmetry explicitly. Since this symmetry must be broken down to $\Urm(1)_{\rm PQ} \times \Urm(1)_Y$, the number of terms in $V_{\rm eb}$ required is $n_D - 1$. 

Being a proper scalar potential, all scalar fields develop VEVs $v_f$ in this basis, where the index $f = u_i, d_i, e_i, S$ is introduced for compactness. Expanding around these VEVs yields,
	\begin{align}\nonumber
		H_{d_i}
			&\supset
				\frac{v_{d_i}}{\sqrt{2}}
				\erm^{i \frac{a_{d_i}}{v_{d_i}}}
				\begin{pmatrix}
				0 \\
				1
				\end{pmatrix}
			\; , 
			&
		H_{u_i}
		    &\supset
				\frac{v_{u_i}}{\sqrt{2}}
				\erm^{i \frac{a_{u_i}}{v_{u_i}}}
				\begin{pmatrix}
					1 \\
					0
				\end{pmatrix}
			\; , \\
		H_{e_i}
			&\supset
				\frac{v_{e_i}}{\sqrt{2}}
				\erm^{i \frac{a_{e_i}}{v_{e_i}}}
				\begin{pmatrix}
				0 \\
				1
				\end{pmatrix}
			\; ,
			&
		S 
			&\supset
				\frac{v_S}{\sqrt{2}}
				\erm^{i \frac{a_S}{v_S}}
			\; ,
	\end{align}
where any angular degrees of freedom not containing the axion are neglected. Each angular mode $a_f$ transforms under a PQ transformation as $a_f \rightarrow a_f + \kappa_f \chi_f v_f$, where the $\chi_f$ denote the PQ charges and the $\kappa_f$ are constants. The corresponding PQ current after spontaneous symmetry breaking is then
	\begin{align}\nonumber
		J_\mu^\PQ 
		\Big|_a
			&\supset
				- \chi_S
					S^\dagger 
					i \partial^\mu
					S
				- \sum_{f \backslash S}
				  \chi_f
					H_f^\dagger 
					i \partial^\mu
					H_f
                + {\rm h.c.}
					\\
			&= 
				\sum_{f}
				\chi_f
				v_f \partial_\mu a_f
			\; .
		\label{Eq:PQCurrent}
	\end{align} 
By requiring $J_\mu^\PQ |_a = v_a \partial_\mu a$ and $a \rightarrow a + \kappa v_a$ under the PQ transformation, the axion field is defined as
	\begin{equation}
		a
			=
				\frac{1}{v_a}
				\sum_{f} \chi_f v_f a_f
			\; ,
			\qquad
		v_a^2 
			= 
				\sum_{f} 
				\chi_f^2 v_f^2 
			\; .
		\label{Eq:DFSZAxion}
	\end{equation}
Thus, in the DFSZ-type models the axion is a linear combination of all scalar angular modes.

With the axion identified, the low-energy theory is constructed in the standard way. By inverting Eq.~(\ref{Eq:DFSZAxion}), the scalar angular modes can be expressed in terms of the axion. Since we are only interested in the terms including the axion, this comes down to the replacement,
    \begin{align}
        \frac{a_f}{v_f}
            \rightarrow 
                \chi_f
                \frac{a}{v_a}
            \; .
    \end{align}
The Lagrangian can then be brought to a more convenient form by a field-dependent chiral redefinition of the fermion fields,
    \begin{equation}
        f
            \rightarrow
                \exp\left(
                    - i \gamma_5 \chi_f
                    \frac{a}{2 v_a}
                \right)
                f
            \; .
    \end{equation}

This redefinition removes the axion from the fermion mass terms, but due to the invariance of the kinetic terms it induces derivative couplings to the fermions. In addition, since in general the PQ current is anomalous with respect to QCD and electromagnetism, anomalous couplings to the gluons and the photons are induced,
    \begin{align}\nonumber
        \delta \mathcal{L}_{\rm anomalous} 
            &=
                N
                \frac{a}{v_a}
                \frac{g_s^2}{16 \pi^2}
                G \Tilde{G}
                + E
                \frac{a}{v_a}
                \frac{g^2}{16 \pi^2}
                F \Tilde{F} \\
            &=
                \frac{a}{f_a}
                \frac{g_s^2}{32 \pi^2}
                G \Tilde{G}
                + \frac{E}{N}
                \frac{a}{f_a}
                \frac{g^2}{32 \pi^2}
                F \Tilde{F}
            \; ,
    \end{align}
where the axion decay constant $f_a \equiv v_a/2N$ is introduced in the second equality. The canonically normalised axion-photon interaction is defined via
    \begin{equation}
        \mathcal{L}_{a\gamma} 
            =
                \frac{1}{4} g_{a\gamma}
                F \tilde{F}
            \; ,
    \end{equation}
so taking into account next-to-leading-order chiral corrections \cite{GrillidiCortona:2015jxo} results in the axion-photon coupling given in Eq.~(\ref{Eq:AxionPhotonC}).

Since in the models under consideration all representations except the PQ charges are known, the ratio between the electromagnetic and colour anomaly coefficients can conveniently be written as \cite{DiLuzio:2017pfr}
    \begin{equation}
        \frac{E}{N}
            =
                \frac{\sum_i \frac{4}{3} \chi_{u_i} + \frac{1}{3} \chi_{d_i} + \chi_{e_j}}
                {\frac{1}{2} \sum_i \chi_{u_i} + \chi_{d_i}}
            =
                \frac{2}{3}
                + 2 \frac{\sum_i \chi_{u_i} + \chi_{e_i}}
                {\sum_i \chi_{u_i} + \chi_{d_i}}
            \; .
        \label{Eq:EoverN}
    \end{equation}
Hence, the determination of the anomaly ratio and thus the axion-photon coupling comes down to the determination of the PQ charges, which we turn to now. For further details regarding the DFSZ axion and an explicit construction of the original DFSZ models, see \cite{DiLuzioAxionLandscape}.

%%%%%%%%%%%%%%%%%%%%%%%%%%%%%%%%%%%%%%%%%%%%%%%%%%%%%%%%
%%%%%%%%%%%%%%%%%%%%%%%%%%%%%%%%%%%%%%%%%%%%%%%%%%%%%%%%
\subsection{The PQ Charges}
\label{Subsec:The_PQ_Charges}
\vs{-5mm}

The key point regarding the PQ charges is that the explicit breaking of the $\Urm(1)^{n_D+1}$ symmetry into $\Urm(1)_{\rm PQ} \times \Urm(1)_Y$ must respect the following conditions \cite{DiLuzioAxionLandscape}:
    \begin{enumerate}
        \item Orthogonality between the PQ current $J_\mu^\PQ$ and the weak hypercharge current $J_\mu^Y$.
        \item Invariance under PQ symmetry.
        \item Well-definiteness of domain wall (DW) number $N_\DW$.
    \end{enumerate}
Consequently, the PQ charges are not arbitrary but interrelated by the $n_D + 1$ relations following from these conditions. Solving the resulting LSE then yields a solution for all PQ charges. 

To begin with, the orthogonality requirement between the PQ current defined in Eq.~(\ref{Eq:PQCurrent}) and the weak hypercharge current $J_\mu^Y|_a = \sum_f Y_f v_f \partial_\mu a_f$ implies
    \begin{equation}
        \sum_f \chi_f Y_f v_f^2 
            = 
                0
            \; .
        \label{Eq:Orthogonality}
    \end{equation}
From this relation one can immediately see that in general the PQ charges are not integer numbers. This can also be concluded by the fact that $\Urm(1)_\PQ$ is not compact.

For the PQ invariance, we divide the $n_D - 1$ terms of $V_{\rm eb}$ into two kinds, namely terms consisting of two doublets and two times the singlet, or terms with four doublets. We denote them symbolically as $HHSS$ and $HHHH$ in the following. We restrict ourselves to renormalisable terms so that higher orders in the scalars do not appear. Since the axion must be rendered invisible, there must be at least one term of the form $HHSS$. The form of the other $n_D - 2$ terms is then in principle free. 

It is crucial that the terms in $V_{\rm eb}$ are chosen such that they give rise to linearly independent conditions. In other words, we require $V_{\rm eb}$ to have enough terms to render the system exactly solvable and not underdetermined nor overdetermined. Underdetermined systems do not explicitly break enough of the $\Urm(1)^{n_D+1}$ symmetry, hence giving rise to undesired massless states. Overdetermined systems, on the other hand, are inconsistent (we come back to systems with linearly dependent terms in Sec.~\ref{Subsec:Multiplicity}).

It should also briefly be mentioned that all resulting sets of PQ charges with $N=0$ do not solve the strong CP problem and should thus be discarded.

With the conditions of orthogonality and PQ invariance, it is reasonable to solve for all PQ charges in terms of $\chi_S$, which is otherwise unconstrained as a singlet. The value of $\chi_S$ is irrelevant for the anomaly ratio since it cancels in the ratio. Hence, for a set of terms chosen in $V_{\rm eb}$ we can express all PQ charges in terms of $\chi_S$ and calculate the anomaly ratio. This is the key message of this subsection.

However, there are quantities in which $\chi_S$ does not cancel. One of these quantities is the DW number. Since there is a potential cosmological problem associated with higher DW numbers that we discuss in Sec.~\ref{Subsec:Selection_Criteria}, it is useful to fix $\chi_S$ as well. In particular, it turns out that in theories where the axion is a linear combination of fields a consistency condition on $\chi_S$ follows from the DW number $N_\DW$ being integer-valued.

In the low-energy regime the QCD anomaly induces a periodic potential to the axion. Let us for illustrative purpose take the potential induced by instantons in the dilute instanton gas approximation \cite{THOOFT1986357, PhysRevD.17.2717}, 
    \begin{equation}
		V(\theta)	
			=
				\Lambda_\QCD^4
				\left[
				1 - \cos \left( 2N
				    \frac{a}{v_a}
				    \right)
				\right]
			\; ,
	\label{Eq:PotDIGA}
	\end{equation}
where we expressed the axion decay constant $f_a =v_a/2N$ in terms of $v_a$. The periodicity of the potential results in discrete vacua and the number of these vacua in a single $2\pi$-loop is the DW number $N_\DW$, which can be read off to be $N_\DW = 2N$. In the language of symmetries, this potential explicitly breaks the original $\Urm(1)_\PQ$ down to the discrete group $\mathbb{Z}_{{N}_\DW}$ under which the axion transforms as $a \rightarrow a + 2 \pi n f_a $ with $n \in \mathbb{Z}$. The DW number is encoded in this transformation and it is given by the $n$ that results in a single loop of circumference $2 \pi v_a$, yielding again $N_\DW = 2N$.

However, there is a caveat in theories where the axion is a linear combination of angular modes $a_f$ \cite{Ernst:2018bib}. There, each angular mode also has a residual cyclic  symmetry from its explicit breaking, namely $a_f \rightarrow a_f + 2\pi n_f v_f$ where $n_f \in \mathbb{Z}$. To take these residual symmetries into account, we apply them on both sides of the first equation in Eq.~(\ref{Eq:DFSZAxion}) and read off the DW number as defined in the previous paragraph,
    \begin{equation}
        N_\DW 
            =
                    2 N 
                    \frac{\sum n_f \chi_f v_f^2}{\sum \chi_f^2 v_f^2}
                    \; .  
        \label{Eq:DWN}
    \end{equation}
For the DW number to be integer, we must demand the fraction in this expression to be integer (as it turns out, this can be chosen to be one without loss of generality). The simplest way is given by $n_f = \chi_f$, which would require the compactness of each $U(1)$ and is thus very restrictive. A less restrictive alternative can be found by plugging in the orthogonality condition into the numerator and the denominator of the fraction to remove one of the $v_f$ and then comparing terms with the same $v_f^2$. Let us for simplicity perform this in DFSZ$_2$, where $f=u,d,S$, and require the fraction to be one. We find
    \begin{align}
        n_S 
            &= 
                \chi_S 
            \; ,
    \label{Eq:IntegerCondition1} \\
        n_u
        +  
        n_d 
            &= 
                \chi_u
                +  
                \chi_d
            = 2\chi_S
        \; ,
    \label{Eq:IntegerCondition2}
    \end{align}
where in the second equality we used the PQ invariance from the unique $V_{\rm eb}$ term, i.e.~$H_u H_d S S$. We see that the residual cyclic symmetries of the underlying angular modes result in the condition $\chi_S \in \mathbb{Z}$ in the minimal DFSZ model.

Repeating this procedure for larger numbers of doublets, we find that Eq.~(\ref{Eq:IntegerCondition1}) is always present and that there are more relations of the type of Eq.~(\ref{Eq:IntegerCondition2}). These relations imply that for the DW number to be integer, $\chi_S$ and certain combinations of PQ charges must be integer. In particular, $2N = \sum_i \chi_{u_i} + \chi_{d_i}$ and $\sum_i \chi_{u_i} + \chi_{e_i}$ are such combinations. The key difference for non-minimal models is that fulfilling all appearing conditions is more restrictive than in the minimal case, for instance requiring the minimal value of $\chi_S$ to be integer and larger than one.

To summarise, we see that in theories where the axion is a linear combination of fields, the DW number can again be written as $N_\DW =2N$ but with the premise that the fraction in Eq.~(\ref{Eq:DWN}) is one. This additional condition comes down to the requirement of $\chi_S$ being integer but not necessarily one. For the sake of the discussion in Sec.~\ref{Subsec:Selection_Criteria} it is useful to fix $\chi_S$ to its minimal possible value. Thus, we conveniently define the DW number as
    \begin{equation}
        N_\DW 
           =
                {\rm min \ positive \ integer}
                \left\{
                    2 N 
                \right\}
            \; 
        \label{Eq:DWN2}
    \end{equation}
and use this definition for the remainder of this work.

%%%%%%%%%%%%%%%%%%%%%%%%%%%%%%%%%%%%%%%%%%%%%%%%%%%%%%%%
%%%%%%%%%%%%%%%%%%%%%%%%%%%%%%%%%%%%%%%%%%%%%%%%%%%%%%%%
\subsection{Multiplicity}
\label{Subsec:Multiplicity}
\vs{-5mm}

The way the PQ charges are fixed, as described in the previous subsection, makes the calculation of all possible anomaly ratios straightforward, at least in principle. However, when it comes to defining a notion of \textit{multiplicity}, further specification is needed because of potential overcounting of models. When constructing models, the standard mantra is to include all possible terms compatible with the given symmetries. If for some reason terms are not included at tree-level, without protection from an underlying symmetry these terms will be generated at higher orders.
%For our purpose this means that models that can be added should not be considered as different. 

Regarding $V_{\rm eb}$ this has the implication that potentials that give rise to the \textit{same} PQ charges should not be considered different since they can simply be added. This can be understood in the language of conditions and LSEs. The construction described in Sec.~\ref{Subsec:The_PQ_Charges} required $n_D-1$ terms in the explicit breaking potential. Less terms would result in undesired Goldstone bosons, while too many independent terms result in overdetermined systems that have $\chi_f=0$ for all $f$ and thus do not solve the strong CP problem. However, one can add more and more terms to the potential that give rise to redundant conditions. These are exactly the potentials that have the same solution of the underlying LSE, i.e.~that have the same PQ charges.

This reasoning also has consequences for the Yukawa sector. The construction we described starts by coupling a single doublet to each right-handed fermion but in principle several doublets can couple to the same right-handed fermion. For this reason we complete the Yukawa sector \textit{a posteriori} for each set of possible PQ charges. For instance, such a completion of the Yukawa sector could look as follows. If we find as a possible solution for some LSE that $\chi_{d1} = \chi_{e1}$ then the Yukawa sector for that solution becomes
    \begin{align} \nonumber
		& y^d_{1j} H^{d_1} \bar{Q}_L^1 d_R^j
		\longrightarrow
		(y^d_{1j} H^{d_1} + \tilde{y}^d_{1j} H^{e_1}) \bar{Q}_L^1 d_R^j \; , \\
		& y^e_{1j} H^{e_1} \bar{E}_L^i e_R^j
		\longrightarrow
		(\tilde{y}^e_{1j} H^{d_1} + y^e_{1j} H^{e_1}) \bar{E}_L^1 e_R^j
			\; .
		\label{Eq:CompleteYukawa}
	\end{align}
This guarantees that all possible Yukawa terms compatible with a given solution are included (such as cross-couplings where for instance up-type doublets couple to down-type fermions). In addition, since every set of PQ charges is unique after adding the potentials, the Yukawa sector with all compatible couplings is uniquely determined and no additional multiplicities need to be taken into account.

Adding the potentials and completing the Yukawa sectors for a specific set of PQ charges specifies one model for the counting of the multiplicity. The last step is then to calculate the anomaly ratio for each model and count its multiplicity, which completes the construction procedure.

%%%%%%%%%%%%%%%%%%%%%%%%%%%%%%%%%%%%%%%%%%%%%%%%%%%%%%%%
%%%%%%%%%%%%%%%%%%%%%%%%%%%%%%%%%%%%%%%%%%%%%%%%%%%%%%%%
\subsection{Selection Criteria}
\label{Subsec:Selection_Criteria}
\vs{-5mm}

With the models specified, the question arises if it is possible to impose (phenomenological) selection criteria in order to extract preferred axion models. 

In the KSVZ-type models, which add additional heavy quarks and one singlet scalar to the SM, all of the selection criteria follow from the presence of the new fermions \cite{DiLuzio:2017pfr}. For instance if the new quarks are too heavy and too long lived, they are subject to strong BBN and CMB bounds. Moreover, since their mass is related to $f_a$, the concrete value of $f_a$ plays an important role. Lastly, the presence of additional quarks dramatically affects the running of the QCD coupling constant, potentially spoiling asymptotic freedom and leading to Landau poles below the Planck scale. All these bounds are not present in the DFSZ case, so we are not discussing them further (see \cite{DiLuzio:2017pfr} or \cite{Plakkot} for a detailed discussion). 

Next, let us briefly discuss the aspects that are present in both types of invisible axion models, starting with the DW problem. As mentioned in Sec.~\ref{Subsec:The_PQ_Charges}, at temperatures of order of the QCD scale $T \sim \Lambda_\QCD$, non-perturbative QCD effects generate an effective potential \cite{QCDINstantonsFiniteTemp}. This potential explicitly breaks the original PQ symmetry down to the discrete group $\mathbb{Z}_{{N}_\DW}$, which is then spontaneously broken by one of the vacua. This leads to the formation of DWs that attach themselves to the cosmic strings (from the spontaneous breaking of the PQ symmetry at $T \sim f_a$) and form string-wall systems. For $N_\DW > 1$, the strings stabilise the DWs so that these would dominate the energy density of the universe --- this is the DW problem \cite{DWProblem} (see \cite{Vilenkin:2000jqa} for a review). Thus, one could impose $N_\DW = 1$ as a selection criterion for axion models.

However, there are several ways to avoid the DW problem. First of all, it is not present when the PQ symmetry is broken during or before inflation since then no DWs form inside our Hubble sphere. In the scenario when the PQ symmetry is broken after inflation, it is also possible that the symmetry is not restored at high $T$, thus avoiding the production of strings and walls \cite{Dvali:1995cc}. Alternatively, by embedding the discrete subgroup into a continuous group, the different vacua become related via symmetry transformations, which results in an effective DW number of unity \cite{LAZARIDES198221}. Because of these known solutions, we do not consider $N_\DW = 1$ to have a sufficient level of generality to represent a necessary selection criterion for our main analysis. Nevertheless, we do calculate the DW number for DFSZ$_3$ to DFSZ$_7$, demonstrate the influence of this selection criterion, and compare with the KSVZ case in Sec.~\ref{Subsec:Experimental_Constraints}.

Furthermore, staying in the same category of aspects that are present for both types of invisible axion models, the presence of additional Higgs doublets alters the running of the electroweak gauge coupling. In particular, the maximal case of $n_D = 9$ seems to improve unification with respect to the SM but the resulting unification scale of $\Lambda_{\rm GUT} \sim 10^{13} \GeV$ leads to unacceptable fast proton decay. For this reason and for the sake of a better comparability with the KSVZ case, we consider improvement of unification not applicable as a selection criterion. In addition, it should also be mentioned that for $n_D \sim 50$ asymptotic freedom is spoiled and a Landau Pole appears below the Planck scale, providing a hard upper limit on the number of doublets \cite{DiLuzio:2017pfr}.

Let us finally turn to an aspect that is only present in the DFSZ-type models, namely the general feature of multi-Higgs doublet models to include flavour-changing-neutral-currents (FCNC). Since FCNCs are subject to strong experimental constraints \cite{Pich:2011nh}, they could in principle severely reduce the number of viable DFSZ-type models. However, similar to the DW problem, there are known ways to avoid these FCNC (see \cite{Ivanov:2017dad} for a review):

\textbf{Natural flavor conservation:} The easiest way to avoid FCNCs is to impose the Weinberg-Glashow-Paschos condition \cite{PhysRevD.15.1958, PhysRevD.15.1966}, which requires all right-handed fermions of a given electric charge to couple to only one of the doublets. Imposing this condition effectively sets several Yukawa couplings to zero, which for $n_D > 3$ results in $n_D - 3$ decoupled Higgs doublets. Hence, for DFSZ-type models as we have defined them in the beginning of this section, natural flavor conservation is only possible for $n_D \leq 3$.

\textbf{Flavour alignment:} A less restrictive possibility is to impose an alignment condition, i.e.~requiring the Yukawa matrices of each right-handed fermion to be proportional to one Yukawa matrix. All Yukawa matrices are then simultaneously diagonalised in the fermion mass eigenbasis, yielding no FCNC at tree-level \cite{Gogberashvili:1991ws, Pich:2009sp, deMedeirosVarzielas:2019dyu}.

\textbf{Mass matrix ansätze:} Another possibility is to take the Yukawa matrices to have a specific texture in flavour space. This allows viable SM mass and mixing phenomenology and sufficient suppression of the tree-level FCNCs \cite{PhysRevD.35.3484}.

Natural flavor conservation and mass matrix ansätze are usually implemented by imposing (discrete) symmetries, which also protect the flavour structure from quantum corrections. However, imposing additional symmetries on the scalar potential spoils the so-called decoupling property of general multi-Higgs doublet models \cite{PhysRevD.103.075026}. This means that the new scalar cannot have arbitrary large masses, resulting in potentially significant deviations from the measured SM couplings. So in order to avoid FCNC using these solutions, it would be necessary to systematically determine which of our models have discrete symmetries that avoid FCNC and at the same time allow for a decoupling limit. Due to the large number of models and the lack of a catalogue of possible symmetries for $n_D > 3$ \cite{Ivanov:2017dad}, such an analysis goes beyond the scope of this work.

On the other hand, flavour alignment is usually assumed without an underlying symmetry protection. While this preserves the decoupling limit of general multi-Higgs doublet models, it leaves the flavour structure vulnerable to quantum corrections. However, due to residual flavour symmetries the induced misalignment is sufficiently small \cite{Penuelas:2017ikk}. 

All in all, in the DFSZ case we find desirable features for specific models but no selection criteria with a sufficient level of generality.

Lastly, we want to mention that in principle it is possible to enlarge the definition of DFSZ-type axions to include more singlets or more than $n_D=9$ doublets, which do not couple to the SM fermions. From the point of view of possible axion-photon couplings, this does not change Eq.~(\ref{Eq:EoverN}), however it allows to obtain very large PQ charges \cite{DiLuzio:2017pfr, Farina:2016tgd}. We do not consider these models in this paper and stick with the more narrow definition of DFSZ models given in the beginning of this section. One could also see this as a kind of selection criterion.

%%%%%%%%%%%%%%%%%%%%%%%%%%%%%%%%%%%%%%%%%%%%%%%%%%%%%%%%
%%%%%%%%%%%%%%%%%%%%%%%%%%%%%%%%%%%%%%%%%%%%%%%%%%%%%%%%
\section{ANOMALY RATIO DISTRIBUTIONS}
\label{sec:ANOMALY_RATIO_DISTRIBUTIONS}
\vs{-5mm}

%%%%%%%%%%%%%%%%%%%%%%%%%%%%%%%%%%%%%%%%%%%%%%%%%%%%%%%%
%%%%%%%%%%%%%%%%%%%%%%%%%%%%%%%%%%%%%%%%%%%%%%%%%%%%%%%%
\subsection{Approach}
\label{Subsec:Approach}
\vs{-5mm}

In the previous section, we reviewed how for the DFSZ-type axions the calculation of the anomaly ratio reduces to fixing a $V_{\rm eb}$ and solving the resulting LSE. Hence, in order to calculate all possible anomaly ratios one has to do exactly that for all possible $V_{\rm eb}$. 

In addition, adding different $V_{\rm eb}$ that give rise to the same set of PQ charges we count how many different sets of charges lead to the same anomaly ratio. This notion of \textit{multiplicity} of each anomaly ratio allows us to allocate a certain probability to each anomaly ratio within the given set of models. We then use this to define lower $|\mathcal{C}_{a\gamma}|$ bounds above which most of the probability mass of DFSZ-type axion models can be found.

In the form of a cooking recipe, our procedure can be summarised by the following steps:
\begin{enumerate}
    \item Specify the Yukawa sector for a fixed $n_D$ by coupling one doublet to each right-handed fermion. This exhausts the maximal freedom regarding the anomaly ratio.
    \item Write down all possible $V_{\rm eb}$ with $n_D - 1$ terms.
    \item Solve all associated LSEs to find all possible sets of PQ charges. Underdetermined systems are discarded.
    \item Add the potentials of all equal PQ charges to get the most general potential associated with a particular solution. This defines one model for the sake of counting the multiplicity.
    \item For each model, complete the Yukawa sector by adding all Yukawa terms compatible with the PQ- and hypercharges.
    \item For each model, calculate the anomaly ratio and count its multiplicity.
\end{enumerate}

We calculate the PQ charges and anomaly ratios numerically using the programming language ``Julia" \cite{julia}. The ``StaticArrays" package \cite{StaticArrays} allows us to compute the extremely large number of LSEs very fast without heap memory allocation. Since it is not relevant to the acquired solutions, we skip step 5 in practice.

%%%%%%%%%%%%%%%%%%%%%%%%%%%%%%%%%%%%%%%%%%%%%%%%%%%%%%%%
%%%%%%%%%%%%%%%%%%%%%%%%%%%%%%%%%%%%%%%%%%%%%%%%%%%%%%%%
\subsection[Example: $n_D = 3$]{Example: $\mathbf{n_D = 3}$}
\label{Subsec:nd=3}
\vs{-5mm}

\begin{table*}
    \centering
    \caption{Resulting PQ conditions from quadrilinears, constructed from corresponding bilinears. The lower triangle ('$-$') is not to be counted because the order of the bilinears does not matter. The terms 'x' are not to be counted because they are Hermitian to a term that has already been counted and the potential by definition has to include all Hermitian conjugated terms. Terms with 'o' produce only trivial conditions. We are left with 9 distinct quadrilinears, which produce 6 unique conditions.}
    \renewcommand{\arraystretch}{1.1}
    %\small
    \begin{tabularx}{\linewidth}{l| Z Z Z | Z Z Z}
    \toprule
                     & $(H_u H_d)$ & $(H_u H_e)$ & $(H_d H_e^\dagger)$ & $(H_u H_d)^\dagger$ & $(H_u H_e)^\dagger$ & $(H_d H_e^\dagger)^\dagger$  \\ \midrule
         $(H_u H_d)$ & $2\chi_u + 2\chi_d=0$ & $2\chi_u + \chi_d + \chi_e=0$ & $\chi_u + 2\chi_d - \chi_e = 0$ & o & $\chi_d - \chi_e = 0$ & $\chi_u + \chi_e = 0$ \\
         $(H_u H_e)$ & $-$ & $2\chi_u + 2\chi_e = 0$ & $\chi_u + \chi_d = 0$ & x & o & $\chi_u -\chi_d + 2\chi_e = 0$\\
         $(H_d H_e^\dagger)$ & $-$ & $-$ & $2\chi_d - 2\chi_e = 0$ & x & x & o  \\ \midrule
         $(H_u H_d)^\dagger$ & $-$ & $-$ & $-$ & x & x & x \\
         $(H_u H_e)^\dagger$ & $-$ & $-$ & $-$ & $-$ & x & x \\
         $(H_d H_e^\dagger)^\dagger$ & $-$ & $-$ & $-$ & $-$ & $-$ & x \\ \bottomrule
    \end{tabularx}
    \label{tab:quads}
\end{table*}

\begin{table*}
    \renewcommand{\arraystretch}{1.1}
    \centering
    \caption{All possible solutions for PQ charges of Higgs doublets in terms of $\chi_S$ (\textbf{top}) and anomaly ratios (\textbf{bottom}) for the $n_D=3$ Yukawa sector under consideration. The potential should not produce the same condition twice ('x'), nor does the order of the conditions matter ('$-$'). 'o'  denotes combinations of conditions that do not have a solution. Infinite solutions arise when $N=0$.}
    \footnotesize
    \begin{tabularx}{\linewidth}{Z| Y Y Y | Y Y Y}
    \toprule
         $[\chi_u, \chi_d, \chi_e]$                         & $\chi_u + \chi_d=2\chi_S$ & $\chi_u + \chi_e=2\chi_S$ & $\chi_d - \chi_e=2\chi_S$ & $-\chi_u - \chi_d=2\chi_S$ & $-\chi_u - \chi_e=2\chi_S$ & $-\chi_d + \chi_e=2\chi_S$  \\ \midrule
        $\chi_u + \chi_d=2\chi_S$  & x                  & $-$ & $-$ & $-$ & $-$ & $-$ \\
        $\chi_u + \chi_e=2\chi_S$  & $[4/3, 2/3, 2/3]$  & x & $-$ & $[0, -2, 2]$ & $-$ & $-$ \\
        $\chi_d - \chi_e=2\chi_S$  & $[2/3, 4/3, -2/3]$ & $[2, 2, 0]$ & x  &  $[-2, 0, -2]$ & $[-2/3, 2/3, -4/3]$ & $-$  \\
        $-\chi_u - \chi_d=2\chi_S$ & o                & $-$ & $-$ & x & $-$ & $-$  \\
        $-\chi_u - \chi_e=2\chi_S$ &  $[0, 2, -2]$     & o & $-$ & $[-4/3, -2/3, -2/3]$ & x & $-$  \\
        $-\chi_d + \chi_e=2\chi_S$ & $[2, 0, 2]$       & $[2/3, -2/3, 4/3]$ & o & $[-2/3, -4/3, 2/3]$& $[-2, -2, 0]$ & x \\ \midrule
        $2\chi_u + 2\chi_d=0$      & o & $[2/3, -2/3, 4/3]$ & $[-2/3, 2/3, -4/3]$ & o & $[-2/3, 2/3, -4/3]$ & $[2/3, -2/3, 4/3]$ \\
        $\chi_u + \chi_d=0$        & o & $[2/3, -2/3, 4/3]$ & $[-2/3, 2/3, -4/3]$ & o & $[-2/3, 2/3, -4/3]$ & $[2/3, -2/3, 4/3]$ \\
        $2\chi_u + 2\chi_e = 0$    & $[2/3, 4/3, -2/3]$ & o & $[2/3, 4/3, -2/3]$ & $[-2/3, -4/3, 2/3]$ & o & $[-2/3, -4/3, 2/3]$ \\
        $\chi_u + \chi_e = 0$      & $[2/3, 4/3, -2/3]$ & o & $[2/3, 4/3, -2/3]$ & $[-2/3, -4/3, 2/3]$ & o & $[-2/3, -4/3, 2/3]$ \\
        $2\chi_d - 2\chi_e = 0$    & $[4/3, 2/3, 2/3]$ & $[4/3, 2/3, 2/3]$ & o & $[-4/3, -2/3, -2/3]$ & $[-4/3, -2/3, -2/3]$ & o \\
        $\chi_d - \chi_e = 0$      & $[4/3, 2/3, 2/3]$ & $[4/3, 2/3, 2/3]$ & o & $[-4/3, -2/3, -2/3]$ & $[-4/3, -2/3, -2/3]$ & o \\
        $2\chi_u + \chi_d + \chi_e=0$ & $[0, 2, -2]$ & $[0, -2, 2]$ & $[0, 1, -1]$ & $[0, -2, 2]$ & $[0, 2, -2]$ & $[0, -1, 1]$  \\
        $\chi_u + 2\chi_d - \chi_e = 0$ & $[2, 0, 2]$ &  $[1, 0, 1]$ & $[-2, 0, -2]$ & $[-2, 0, -2]$ &  $[-1, 0, -1]$ & $[2, 0, 2]$ \\
        $\chi_u -\chi_d + 2\chi_e = 0$ &  $[1, 1, 0]$ &  $[2, 2, 0]$ & $[2, 2, 0]$ &  $[-1, -1, 0]$ &  $[-2, -2, 0]$ & $[-2, -2, 0]$ \\ \bottomrule
    \end{tabularx}
        \vskip 0.5cm
        \small
    \begin{tabularx}{\linewidth}{Z| Y Y Y | Y Y Y}%{c|c c c | c c c}
    \toprule
        $E/N$                   & $\chi_u + \chi_d=2\chi_S$ & $\chi_u + \chi_e=2\chi_S$ & $\chi_d - \chi_e=2\chi_S$ & $-\chi_u - \chi_d=2\chi_S$ & $-\chi_u - \chi_e=2\chi_S$ & $-\chi_d + \chi_e=2\chi_S$  \\ \midrule
        $\chi_u + \chi_d=2\chi_S$ & x & $-$ & $-$ & $-$ & $-$ & $-$ \\
        $\chi_u + \chi_e=2\chi_S$ & $8/3$ & x & $-$ & $-4/3$ & $-$ & $-$ \\
        $\chi_d - \chi_e=2\chi_S$ & $2/3$ & $5/3$ & x  &  $14/3$ & $\infty$ & $-$  \\
        $-\chi_u - \chi_d=2\chi_S$ & o & $-$ & $-$ & x & $-$ & $-$  \\
        $-\chi_u - \chi_e=2\chi_S$ & $-4/3$ & o & $-$ & $8/3$ & x & $-$  \\ 
        $-\chi_d + \chi_e=2\chi_S$ & $14/3$ & $\infty$ & o & $2/3$ & $5/3$ & x \\ \midrule
        $2\chi_u + 2\chi_d=0$  & o & $\infty$ & $\infty$ & o & $\infty$ & $\infty$ \\
        $\chi_u + \chi_d=0$    & o & $\infty$ & $\infty$ & o & $\infty$ & $\infty$ \\
        $2\chi_u + 2\chi_e = 0$& $2/3$ & o & $2/3$ & $2/3$ & o & $2/3$ \\
        $\chi_u + \chi_e = 0$  & $2/3$ & o & $2/3$ & $2/3$ & o & $2/3$ \\
        $2\chi_d - 2\chi_e = 0$& $8/3$ & $8/3$ & o & $8/3$ & $8/3$ & o \\
        $\chi_d - \chi_e = 0$  & $8/3$ & $8/3$ & o & $8/3$ & $8/3$ & o \\
        $2\chi_u + \chi_d + \chi_e=0$ & $-4/3$ & $-4/3$ & $-4/3$ & $-4/3$ & $-4/3$ & $-4/3$ \\
        $\chi_u + 2\chi_d - \chi_e = 0$ & $14/3$ & $14/3$ & $14/3$ & $14/3$ & $14/3$ & $14/3$  \\
        $\chi_u -\chi_d + 2\chi_e = 0$ &  $5/3$ & $5/3$ & $5/3$ &  $5/3$ & $5/3$ & $5/3$ \\ \bottomrule
    \end{tabularx}
    \normalsize
    \label{tab:terms}
\end{table*}

Let us demonstrate our approach in the example of DFSZ$_3$ with the Weinberg-Glashow-Paschos condition imposed, i.e.,~with one Higgs doublet per type of fermion. In this example there are three possible bilinears, namely $(H_u H_d)$, $(H_u H_e)$ and $(H_d H_e^\dagger)$, together with their complex conjugates. Each bilinear can either be coupled to the singlet, which results in 6 different terms of the form $HHSS$, or to another bilinear, which results in 36 different quadrilinears of the form $HHHH$. For the latter case, removing terms that are related by Hermitian conjugation and terms that result in no condition reduces the number to 9 (see Tab.~\ref{tab:quads}). For $n_D=3$, the breaking potential consists of either one $HHSS$ and one $HHHH$ term or two $HHSS$ terms. For the former, there are \textit{a priori} $54$ possibilities, and for the latter $15$, totaling to $69$ possibilities for $V_{\rm eb}$ (see Tab.~\ref{tab:terms}).

The resulting $3 \times 3$ LSEs consist of the orthogonality relation, $\chi_u v_u^2 - \chi_d v_d^2 - \chi_e v_e^2= 0$, and the two conditions coming from the potential. Solving the LSEs yields the PQ charges in terms of $\chi_S$, which is then fixed by the well-definiteness of the DW number. We can do the following two simplifications for the purpose of calculating the anomaly ratio. First, we can set all VEVs equal to one because $E$ and $N$ are independent of them, and secondly, we can leave $\chi_S$ unfixed because it cancels in the anomaly ratio after expressing all PQ charges in terms of $\chi_S$. 

Of the $69$ minimal potentials found, many have no or degenerate solutions. For example, potentials including a bilinear and its Hermitian conjugate at the same time do not have a solution and the nine quadrilinears only give six unique conditions for PQ charges. A summary of all solutions can be found in Tab.~\ref{tab:terms} (top). In total, this leaves only $16$ different solutions for the doublet charges, for each of which we have to add all the terms to the potential that give rise to this set of PQ charges.

The Yukawa sector in this example does not need any completion since it is already fixed by the Weinberg-Glashow-Paschos condition. Hence, it merely remains to plug into Eq.~(\ref{Eq:EoverN}) the different sets of PQ charges, which yields the following possible anomaly ratios (see Tab.~\ref{tab:terms}, bottom)
    \begin{equation}
        {\rm DFSZ}_3:
        \frac{E}{N}
            =
                -\frac{4}{3} \; ,
                \frac{2}{3} \; ,
                \frac{5}{3} \; ,
                \frac{8}{3} \; ,
                \frac{14}{3} \; .
    \end{equation}
Counting the multiplicity, we find that $2/3$ and $8/3$ each appear $2$ times with four terms in the potential each, and $-4/3$, $5/3$ as well as $14/3$ each appear $4$ times with three or two terms in the potential each. A visualisation of this result together with all other $n_D$ values can be found in Fig.~\ref{fig:pdfn3to9}. For a summary of important statistics, see Tab.~\ref{tab:catalogue}.

It turns out useful in the following to introduce a compact notation that encodes which doublet couples to which of the nine fermions. For this, we assign to the nine fermions a position in a nine-dimensional row vector with square brackets,
    \begin{align} \nonumber
        & \ \, u \ \ \, c \ \ \, \, t \ \ \, \, d \ \ \, s \ \ \, b \ \ \, e \ \ \, \mu \ \ \, \tau \\  \nonumber
        & \ \, u_1 \, \, u_2 \, \, u_3 \ \, d_1 \, \, d_2 \, \, d_3 \, \, e_1 \, \, \, e_2 \, \, e_3 \\
        & [\ \cdot \ , \ \cdot \ , \ \cdot \ , \ \cdot \ , \ \cdot \ , \ \cdot \ , \ \cdot \ , \ \cdot \ , \ \cdot \ ]
                \; ,
    \end{align}
and write the subscript of the doublets that couple to a certain fermion to the corresponding position. If one doublet couples to multiple fermions, we use the first subscript in the order presented above. For more comprehensive notation we use fermion type (up-, down-, or lepton-type, short $u$, $d$, or $e$) and generation ($1$ to $3$). For DFSZ$_9$, this row vector would be $[u1,u2,u3,d1,d2,d3,e1,e2,e3]$ while for the original DFSZ$_2$-I model it would be $[u1,u1,u1,d1,d1,d1,d1,d1,d1]$.

%%%%%%%%%%%%%%%%%%%%%%%%%%%%%%%%%%%%%%%%%%%%%%%%%%%%%%%%
%%%%%%%%%%%%%%%%%%%%%%%%%%%%%%%%%%%%%%%%%%%%%%%%%%%%%%%%
\subsection{Choices for a Statistical Interpretation}
\label{subsec:choices_for_a_statistical_interpretation}
\vs{-5mm}

We are considering many different solutions for the Higgs charges. In Sec.~\ref{Subsec:nd=3}, we just counted the number of models leading to specific anomaly ratios, but in the end we want to translate a catalogue of models with specific $E/N$ values to a probability distribution of anomaly ratios. To achieve this, we require relative probabilities of the solutions, which are subject to some sort of theoretical prior belief. This belief manifests itself in multiple decisions about:

\begin{itemize}
    \item \textbf{The concept of multiplicity} as outlined in Sec.~\ref{Subsec:Multiplicity}.
    \item \textbf{The relative probability of different Yukawa sectors given a specific $\mathbf{n_D}$.} 
    
    A reasonable choice is to demand all solutions with a given $n_D$ to be equally probable. The same applies to different Yukawa sectors. Unfortunately, both cannot be true at the same time because different Yukawa sectors can lead to different amounts of possible solutions. We take the approach of requiring solutions to be equal (given equal multiplicity and same $n_D$). This also implies not applying any ``beauty" arguments for Yukawa sectors, e.g.~in favour of coupling patterns that are equal for different fermion types.
    
    \item \textbf{The relative probabilities of different $\mathbf{n_D}$.}
    
    For our total anomaly ratio distribution, we treat the probability of all $n_D$ values $2 \leq n_D \leq 9$ as equal. This implies at the same time that we consider any single solution for e.g.~DFSZ$_3$ (of which there are 16) much more probable than any single solution for e.g.~DFSZ$_5$ (of which there are $9.7 \times 10^4$). One could also consider it reasonable to additionally penalise models with higher $n_D$, enhance the probability of models satisfying symmetry arguments (DFSZ$_3$, one Higgs doublet per fermion-type or DFSZ$_9$, one Higgs per right-handed fermion) or consider all charge solutions equally probable. In the latter case, the final histogram would most probably be completely dominated by DFSZ$_9$ due to the much larger amount of unique solutions.
\end{itemize}

The arguments above all imply a probabilistic approach to model selection, i.e.~nature ``selects" one of the possible realisations at random. This notion itself may be subject to critique, but in absence of any decisive underlying physical argument singling out any specific model, we deem it to be satisfactory. In Sec.~\ref{Subsec:Selection_Criteria}, we outline theoretical arguments that might challenge this view. 

Even under the assumption of probabilistic model selection we acknowledge that any of these choices is to some extent a matter of taste. For this reason it is important to us to provide the raw catalogues and generating code as a supplement to this paper, so the reader is not dependent on our choice.

%%%%%%%%%%%%%%%%%%%%%%%%%%%%%%%%%%%%%%%%%%%%%%%%%%%%%%%%
%%%%%%%%%%%%%%%%%%%%%%%%%%%%%%%%%%%%%%%%%%%%%%%%%%%%%%%%
\subsection[Results for $n_D=4-7$]{Results for $\mathbf{n_D=4-7}$}
\label{Subsec:nD4-7}
\vs{-5mm}

\begin{figure}
    \centering
    \includegraphics[width=0.4\textwidth]{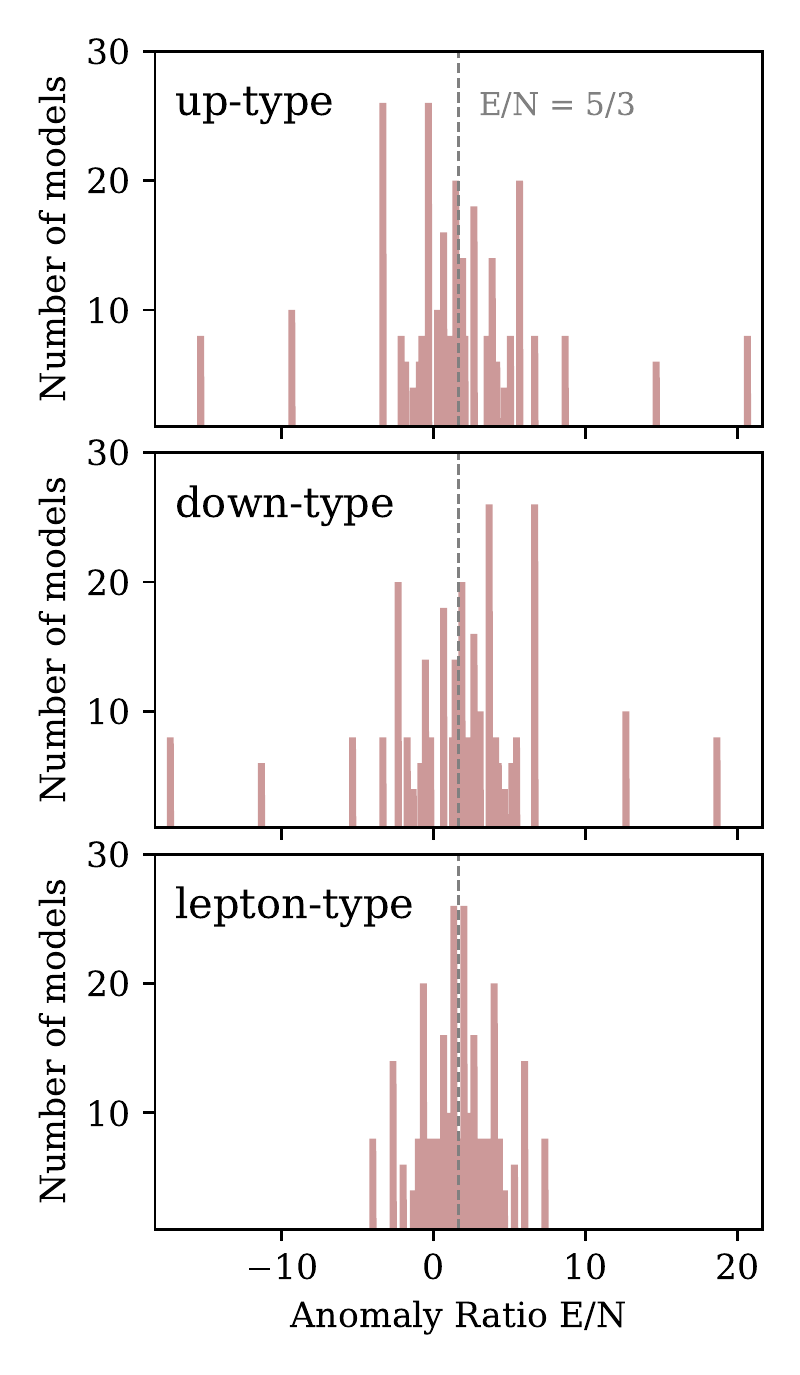}
    \caption{Anomaly ratio distributions for DFSZ-type models with 4 Higgs doublets. Two Higgs couple to the fermions specified in the panels with the other two Higgs covering the remaining two fermion types invariant with respect to fermion generation. For example, the Yukawa sectors $[u1,u1,u3,d1,d1,d1,e1,e1,e1]$, $[u1,u2,u1,d1,d1,d1,e1,e1,e1]$, and $[u1,u2,u2,d1,d1,d1,e1,\\e1,e1]$ are all equivalent and have anomaly ratio distributions as shown in the top panel. Note that the up-type and down-type cases are mirrored around 5/3.}
    \label{fig:n4overview}
\end{figure}

Having presented our assumptions leading to a statistical treatment explicitly, we can now proceed to higher numbers of Higgs doublets, for which we investigate multiple different Yukawa sectors. First, we stick to DFSZ$_4$ to DFSZ$_7$ because for these models we are able to calculate all possible solutions explicitly.

Fig.~\ref{fig:n4overview} presents an overview over the anomaly ratio distributions for DFSZ$_4$ models grouped by the different Yukawa sectors. Each of the histograms shows all models of the specified coupling with the explicitly symmetry breaking potential $V_{\rm eb}$ consisting of $k \geq 1$ $HHSS$- and $3-k$ $HHHH$-terms. The result does not depend on fermion generation since the construction of the Higgs charges as well as Eq.~(\ref{Eq:EoverN}) treat all generations equally. Yukawa sectors with special coupling to a lepton have histograms symmetric around $5/3$, while the histograms for up- and down-type special couplings are mirrored around $5/3$.

The reason for this is a symmetry in our construction as well as in Eq.~(\ref{Eq:EoverN}): For every $n_D$, since we consider all possible Yukawa sectors as outlined above, every solution has a corresponding one with 
    \begin{equation}
        \label{Eq:trafo}
        \begin{aligned}
            \chi_{\tilde{u}_i} \rightarrow& - \chi_{d_i}\\
            \chi_{\tilde{d}_i} \rightarrow& - \chi_{u_i} \,. 
        \end{aligned}
    \end{equation}
This is due to up-type and down-type quarks being treated equally in the construction except for the sign of their hypercharges. In the example above, all solutions for the Yukawa sector $[u1,u1,u3,d1,d1,d1,e1,e1,e1]$ have a corresponding solution in the Yukawa sector $[u1,u1,u1,d1,d1,d3,e1,e1,e1]$ under the above mentioned transformation. Solutions that relate via Eq.~(\ref{Eq:trafo}) can easily be seen to have anomaly ratios relating by 
    \begin{equation}
        \widetilde{\frac{E}{N}} \rightarrow \frac{10}{3} - \frac{E}{N}\,,
    \end{equation}
which is a mirror symmetry around $\frac{5}{3}$.

If we add up all nine histograms of Fig.~\ref{fig:n4overview}, i.e.,~do not treat any Yukawa sector preferentially, we obtain the distribution shown in Fig.~\ref{fig:pdfn3to9} (second row, left). Due to the symmetries of the nine contributing Yukawa sectors, the distribution is symmetric around $5/3$ as well. The biggest number of models coincides with the two possible values for the DFSZ$_2$ model: $2/3$ and $8/3$. Both of these statements are true for $n_D \in [4,7]$, as Fig.~\ref{fig:pdfn3to9} shows (second row, third row left). 

\begin{figure}[t]
    \centering
        \includegraphics[width=0.498\textwidth]{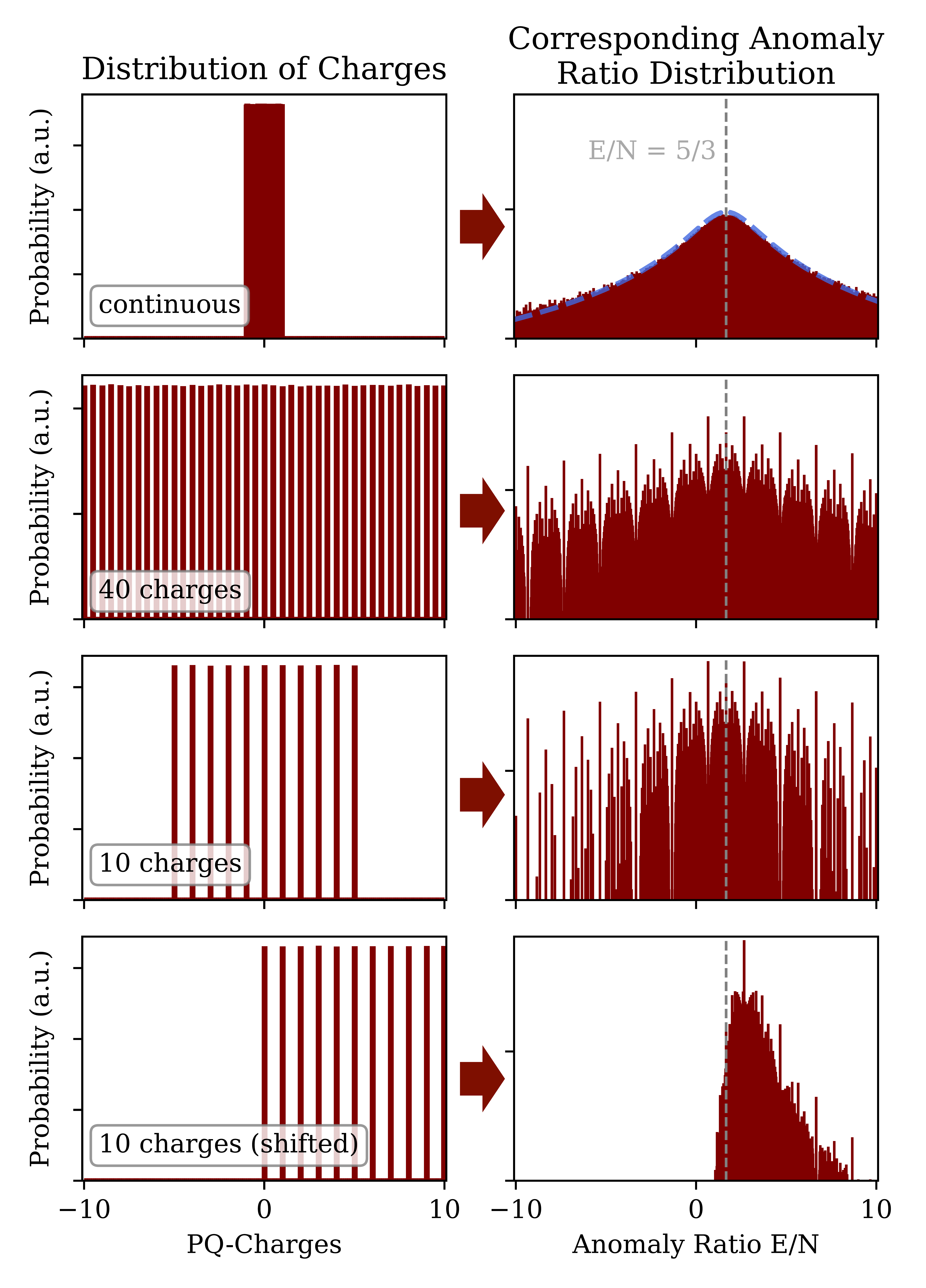}
    \caption{Influence of drawing charges from different distributions on the resulting anomaly ratio distribution, using Eq.~(\ref{Eq:EoverN}). More unique charges lead to a smoother anomaly ratio distribution, irrespective of their distribution. Charge distributions centred around $0$ produce anomaly ratio distributions centred around $5/3$. The dashed blue line in the top right panel denotes the fit presented in Eq.~(\ref{Eq:EoNfit}).}
    \label{fig:EoNtest}
\end{figure}

With increasing $n_D$, we find an increasing number of unique anomaly ratios and more extreme $E/N$ values. Anomaly ratios $E/N = 5/3 + k$ with $k \in \mathbb{Z}$ are highly favoured for $n_D \geq 5$ compared to other $E/N$ values, especially for odd $k$. We see this very characteristic, peaked spectrum evolving: $E/N$ values with high probability tend to have their probabilities shrink with increasing $n_D$, whereas low probability $E/N$ values behave in the opposite manner. In Fig.~\ref{fig:pdfn3to9}, one can most easily see this evolution at big anomaly ratios $E/N \gtrsim 10$. 

Let us try to understand this trend from a purely mathematical perspective:
    \begin{equation*}
        \frac{E}{N}
            =
                \frac{2}{3}
                + 2 \frac{\sum_i \chi_{u_i} + \chi_{e_i}}
                {\sum_i \chi_{u_i} + \chi_{d_i}}
            \; .
    \end{equation*}
is a function with nine variables, the values of each of which can be thought of as being drawn from a specific distribution. In Fig.~\ref{fig:EoNtest}, we show the effect of using different distributions for the variables on the outcome of the function. A continuous, flat charge distribution of arbitrary width produces a smooth, fat-tailed $E/N$ distribution. If the median of the charges is $0$, the median of the distribution is at $5/3$ (Fig.~\ref{fig:EoNtest}, top three rows). Allowing only positive values for the charges shifts the distribution to higher values, with a median of $8/3$ and makes $E/N < 0$ impossible (Fig.~\ref{fig:EoNtest} bottom row). The fewer distinct input values for the charges are used, the more peaked the anomaly ratio structure becomes, i.e.,~anomaly ratios with high relative probability see their likelihood increased and vice versa. This also leads to fewer possible unique $E/N$-values.

The continuum limit with its vanishing skewness and positive kurtosis can be approximated in analytic form via a Pearson type VII distribution \cite{Pearson1916},
\begin{equation}
\label{Eq:EoNfit}
    p\left(\frac{E}{N}\right) = \frac{1}{\alpha \,\mathrm{B}(m-\frac{1}{2},\frac{1}{2})} \left[ 1+ \left(\frac{\frac{E}{N}-\lambda}{\alpha} \right)^2 \right]^{-m}, 
\end{equation}
with reasonable fit parameters $\lambda = 5/3$, $\alpha = 7/4$, and $m=1$, and Beta function $\mathrm{B}$ with $\mathrm{B}(1/2,1/2)=\pi$.

Following these insights from a mathematical perspective it can be understood that the histograms for larger $n_D$ should be smoother, considering that there are more unique solutions (Tab.~\ref{tab:catalogue}). Note, however, that this effect neglects the influence of choosing different probabilities for different solutions. Non-uniform probabilities reduce the effective number of different solutions.\footnote{Just think of the extreme case of say, a charge distribution with 100 unique solutions, in which 10 solutions are $1000\times$ more probable than the other 90. The resulting $E/N$ distribution will behave more as if it came only from 10 unique charges than as if it had 100.} Using our approach of adding all possible potential terms for one solution of charges leads to more comparable probabilities for the charges than if we had separately considered all potentials with the minimal amount of terms to fix the PQ charges (minimal potentials). Therefore the effect of non-uniform charge probabilities is clearly subdominant for DFSZ$_5$ to DFSZ$_7$. We expect this to still be the case even for DFSZ$_8$ and DFSZ$_9$.

%%%%%%%%%%%%%%%%%%%%%%%%%%%%%%%%%%%%%%%%%%%%%%%%%%%%%%%%
%%%%%%%%%%%%%%%%%%%%%%%%%%%%%%%%%%%%%%%%%%%%%%%%%%%%%%%%
\subsection[Extrapolation to $n_D > 7$]{Extrapolation to $\mathbf{n_D > 7}$}
\label{subsec:n>7}
\vs{-5mm}

While our procedure in principle works for any number of doublets, for larger $n_D$ it requires solving an extremely large number of LSEs. In order to see how many, let us estimate the number of all possible terms for step 2 with an arbitrary $n_D$. Since the number of possible bilinears is $n_B = \binom{n_D}{2}$ plus their Hermitian conjugate, there are $2n_B$ terms of the form $HHSS$. Regarding the quadrilinears this results in $(2n_B)^2$ possible terms. Written as a matrix, this yields
    \begin{equation}
        \bordermatrix{     
                    & HH     & (HH)^\dagger      \cr
            HH     & A     & B      \cr
            (HH)^\dagger     & C     & D      \cr
            }   
            \; ,
    \end{equation}
where $A$ denotes the submatrix formed by all terms of the form $HHHH$, $B$ by $HH(HH)^\dagger$, and so on. However, as in the DFSZ$_3$, example there are several equal terms in this matrix that should not be counted. First of all, the whole matrix is symmetric. Secondly, since Hermitian conjugated terms are equal, $D$ is completely redundant with respect to $A$. Lastly, $B$ is anti-symmetric, so that the number reduces to $n_B^2$ quadrilinears. 

\begin{figure*}
    \centering
    \includegraphics[width=0.86\textwidth]{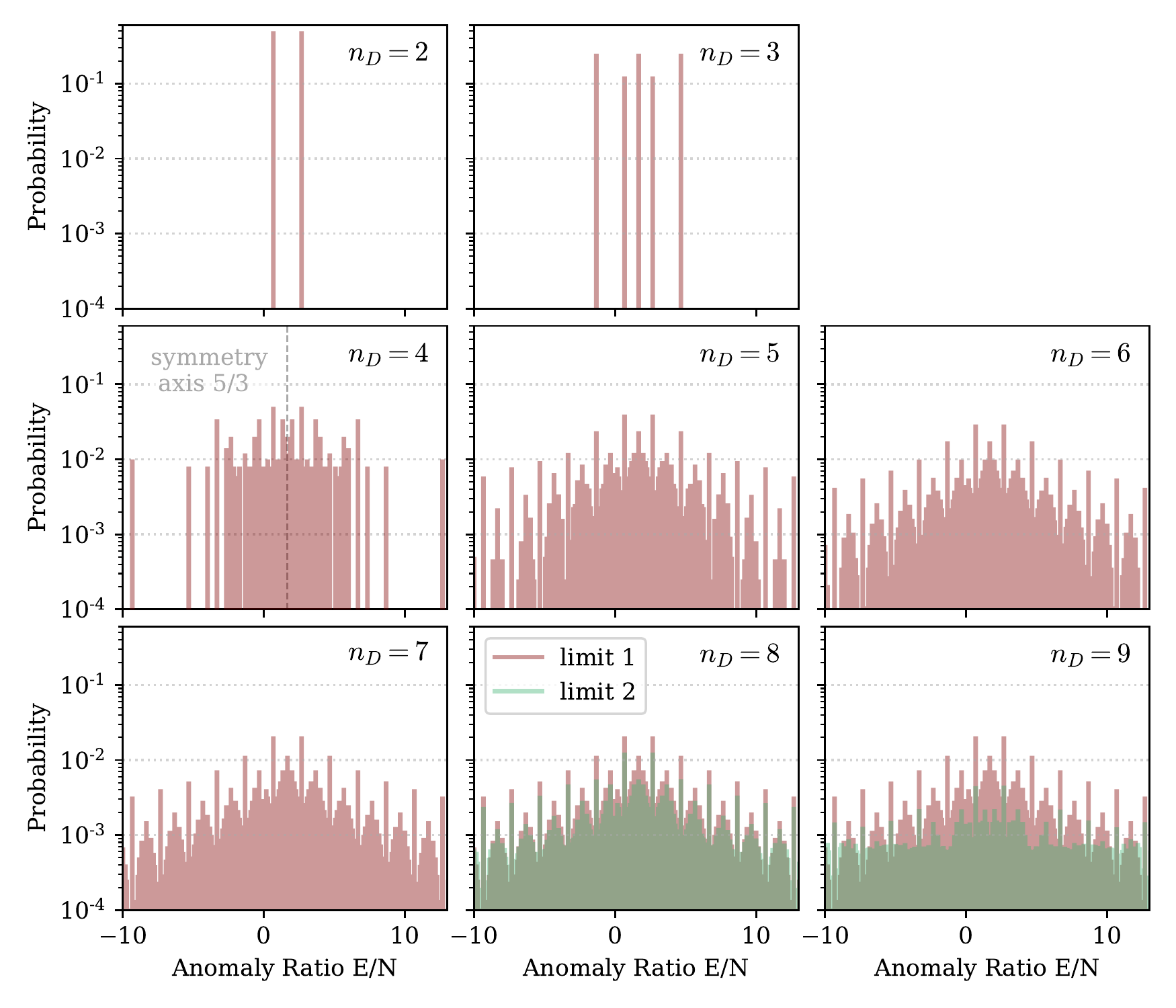}
    \caption{Anomaly ratio distributions for different numbers of Higgs doublets. All histograms are symmetric around $5/3$. $n_D \geq 5$ display a characteristic peaked structure, which becomes smoother with increasing $n_D$. DFSZ$_8$ and DFSZ$_9$ could not be fully calculated, the two semi-transparent colours denote the two estimates as discussed in the text. Note that limit 2 only slightly exceeds limit 1 at big absolute anomaly ratios for $n_D=8$ as well as $n_D=9$.}
    \label{fig:pdfn3to9}
\end{figure*}

\begin{table*}
  \caption{Important statistics of DFSZ-type models broken down by number of Higgs doublets $n_D$. We include information on the model with maximal photon coupling $\widehat{E/N}$ from Eq.~(\ref{eq:maxEoN}) and the percentage of models that have minimal photon coupling (photophobic, $|E/N - 1.92| < 0.04$). 'x' denotes values that could not be estimated.}
  \begin{threeparttable}
\begin{tabularx}{0.758\linewidth}{c | c| c| c |c | c | c} 
 \toprule
 $n_D$ & \#$V_{\rm eb}$ & Unique solutions & Unique $E/N$s & $\widehat{E/N}$ & $\%$ Photophobic & $\%$ $N_\DW = 1$ \\ \midrule
 2 & 2 & 2 & 2 & 2/3 & 0 & 0 \\
 3 & 54 & 16 & 5 & $-4/3$ & 0 & 0 \\
 4 & 52614 & 996 & 83 & $-52/3$ & $1.4$ & $6.00$ \\
  5 & $6.65 \times 10^7$ & $9.7 \times 10^4$ & 432  & $-112/3$ & $1.52$ & $6.64$ \\
  $6^1$ & $\lesssim 4 \times 10^9$ & $> 2.19 \times 10^6$ & 1680  & $-238/3$ & $1.37$ & $5.83$ \\
  $7^1$ & $ \lesssim 7 \times 10^{12}$ & x & 6256 & $-466/3$ & $1.39$ & $5.19$ \\
  $8^2$ & $ \lesssim 2 \times 10^{16}$ & x & $>11617$ & $< -628/3$ & x & x \\
 $9^2$ & $ \lesssim 1 \times 10^{20}$ & x & $\gg 14122$ & $< -1216/3$ & x & x \\ \bottomrule
\end{tabularx}
    \begin{tablenotes}
      \small
      \item $^1$For $n_D \geq 6$, \#$V_{\rm eb}$ and ``unique solutions" are estimates. Number of minimal potentials calculated via Eq.~(\ref{eq:Ntot}), many of which will be unphysical and not produce valid solutions for PQ charges. ``unique solutions" gives the number of solution found in sample, for which data exists.
      \item $^2$For $n_D \geq 8$, we did not calculate all possible models, therefore we have no exact value neither for the number of unique $E/N$, nor for the percentage of photophobic models or models with $N_\mathrm{DW}=1$. $\widehat{E/N}$ was estimated as shown in Sec.~\ref{subsec:maxgag}.
    \end{tablenotes}
  \end{threeparttable}
  \label{tab:catalogue}
\end{table*}

From the set of all terms, we need to pick $n_D - 1$ terms where at least one must be of the form $HHSS$. Hence we can pick between 1 and $n_D-1$ terms of the form $HHSS$, then fill up with $HHHH$ terms, and repeat this for all possible amounts of $HHSS$ terms (ignoring equivalences in the case of multiple $HHSS$ terms). The total number of possible $V_{\rm eb}$ can then be estimated by 
    \begin{equation}
        N_{\rm tot}(n_D)
            \sim  
                \sum_{j=0}^{n_D-2} \binom{2 n_B}{1 + j} \binom{n_B^2}{n_D-2 - j}
            \; ,
        \label{eq:Ntot}
    \end{equation}
which at the same time is the number of LSEs that needs to be solved.

In principle, we can again perform the simplifications used for the DFSZ$_3$ example, namely setting all VEVs to one and not fixing $\chi_S$, but regardless of these simplifications the computation time rises exponentially with $n_D$. While $N_{\rm tot}(n_D = 3) = 69$ is easily manageable, for e.g.~$n_D = 8$ the number of possibilities becomes $N_{\rm tot}(n_D = 8) \approx 2 \cdot 10^{16}$. Thus, computing requirements for solving all LSEs beyond DFSZ$_7$ are prohibitive.

An easy solution to the computationally prohibitive number of LSEs would be to sample the (minimal) potentials. However, due to step 4 in our approach, this is not possible without introducing a bias: A multitude of minimal potentials can lead to the same solution. In our approach, all of them belong to the same model, which for this reason has a very long potential and is likely to be found by any sampling algorithm. On the opposite side, there are also models that can just be found with one or two minimal potentials. Sampling in the space of minimal potentials therefore leads to biased sampling in the space of models.

An alternative estimation for the DFSZ$_8$ and DFSZ$_9$ distributions can come from the following considerations. If a large enough number of theories is considered, Fig.~\ref{fig:EoNtest} (top, right) and Fig.~\ref{fig:EoNtest} (third row, right) can be viewed as extremal cases for the anomaly ratio distribution. ``Extremal" in this context should not be understood in terms of an upper or lower limit on individual $E/N$ bins; after all, we are considering (normalised) probability measures. Rather, Fig.~\ref{fig:EoNtest} (top row, right) is very smooth, whereas Fig.~\ref{fig:EoNtest} (third row, right) is very peaked. Before applying it, let us quantify this criterion by looking at the cumulative sum of anomaly ratios below a specific value. Similarly to the two sample Kolmogorov-Smirnov-test, we define smoothness of an anomaly ratio distribution $f(E/N)$ as
\begin{equation}
\label{Eq:smoothness}
    \mathrm{max}_{x} \left|\sum_{E/N < x} f\left( E/N \right) - \sum_{E/N < x} c\left( E/N \right) \right|,
\end{equation}
where $c(E/N)$ represents the continuous distribution as shown in Fig.~\ref{fig:EoNtest} (top, right). Eq.~(\ref{Eq:smoothness}) defines the maximum of the difference for all anomaly ratios in the cumulative sum of the distribution compared to the case of continuous charges as a possible metric for this task. In Sec.~\ref{Subsec:gag_cdf} we will see the close connection of this metric to the relevant observable. The metric runs from one to zero (by construction for the continuous distribution). For DFSZ$_3$, the value is $17\%$, for DFSZ$_4$ already $5.7\%$, and down to $1.4\%$ for DFSZ$_7$. 

We want to be able to roughly constrain the smoothness of the DFSZ$_8$ and DFSZ$_9$ anomaly ratio distributions. From our results for DFSZ$_3$ to DFSZ$_7$, we saw that the higher the number of doublets, the smoother the anomaly ratio distribution becomes. From investigations of the biased sampling for $n_D=6$ and $n_D=7$, where the true distributions were available, we see that sampling leads to less smooth distributions. This means that the distribution for $n_D=8$ or $n_D=9$ is expected to be smoother than their respective sampled distribution and the $n_D=7$ distribution. In Fig.~\ref{fig:pdfn3to9}, we use $n_D=7$ as one estimate, denoted as ``limit 1".

The other estimate, overestimating the smoothness, can come from the observation that the difference in smoothness of the distributions is smaller between DFSZ$_{6}$ and DFSZ$_7$ than between DFSZ$_{5}$ and DFSZ$_6$. Extrapolating the histograms beyond $n_D = 7$ using the difference of the distributions of DFSZ$_{6}$ and DFSZ$_7$ should therefore yield anomaly ratio distributions which are smoother than our actual expectation. In Fig.~\ref{fig:pdfn3to9}, we subtract the difference once to reach the estimate for $n_D=8$ and twice for $n_D=9$, denoted as ``limit 2". For the metric  described by Eq.~(\ref{Eq:smoothness}), we find $0.73\%$ and $0.71\%$ for DFSZ$_8$ and DFSZ$_9$, respectively. 

The approach presented here should not be viewed as presenting hard limits for the anomaly ratio distributions for eight or nine Higgs doublets, but rather a rough estimate. The difference in probability in Fig.~\ref{fig:pdfn3to9} looks substantial only due to the logarithmic axis. Both estimates are much closer to the continuous case of Fig.~\ref{fig:EoNtest} (top, right) than to the peaked one of Fig.~\ref{fig:EoNtest} (third row, right) in the sense that only very little of their probability mass lies at unique $E/N$ values and rather in a continuum.

\subsection[Constructing extreme $|g_{a\gamma}|$]{Constructing extreme $\mathbf{|g_{a\gamma}|}$}
\label{subsec:maxgag}

Another problem that arises by sampling potentials as described in the previous paragraphs is that it is very unlikely to find the anomaly ratio corresponding to the maximum axion-photon coupling, which we denote as 
    \begin{equation}
        \widehat{E/N} 
            = 
                \mathrm{argmax}_{E/N} (|E/N - 1.92|)
            \; .
    \label{eq:maxEoN}
    \end{equation} 
This anomaly ratio, however, is very useful for constraining the region of DFSZ-type models. For this reason we give a procedure on how to construct an estimate for it. Before turning to this procedure though, let us note that due to the symmetry around $E/N=5/3$, in absence of selection criteria, $\widehat{E/N}$ is not given by the largest possible anomaly ratio but the smallest.

The procedure is based on observations of the LSEs that led to $\widehat{E/N}$ for the smaller numbers of doublets. There, we found that any of the LSEs leading to $\widehat{E/N}$ of DFSZ$_4$ can be extended to an LSE leading to $\widehat{E/N}$ for DFSZ$_5$. The same behaviour can be seen from DFSZ$_5$ to DFSZ$_6$ and in a slightly different form from DFSZ$_3$ to DFSZ$_4$. We do not have a rigorous mathematical reason why this is the case, so applying it to larger $n_D$ is more of an educated guess. However, it turns out to give extreme anomaly ratios, so we use it to systematically estimate $\widehat{E/N}$.

The procedure goes as follows. First, we take all LSEs that lead to $\widehat{E/N}$ for a number of doublets where all solutions are known, say $n_D=6$. Secondly, we add one additional Higgs doublets by specifying the Yukawa sector for the new doublet. Thirdly, we adjust the orthogonality relation appearing in all LSEs depending on what type of doublet is added. Then, we add one additional relation to the LSEs, solve them and calculate the anomaly ratio. After that we repeat this for every possible relation and every possible Yukawa sector. Finally, we extract the LSEs with the smallest anomaly ratio.

This results in highly negative anomaly ratios. However, we found for DFSZ$_9$ that taking the resulting LSEs and systematically exchanging one (or more if the runtime is acceptable) of the relations, new LSEs are found that give even smaller anomaly ratios. In DFSZ$_9$ for instance, the smallest anomaly ratio we construct in this way is $E/N = -1216/3$, and it is generated by the terms
    \begin{align}\nonumber
        &(H_{d_2}^\dagger H_{e_1}) (H_{d_2}^\dagger H_{d_1}) \; , \
        (H_{u_1} H_{d_1}) (H_{u_1} H_{d_2}) \; , \\ \nonumber
        &(H_{u_3}^\dagger H_{u_1}) (H_{u_3}^\dagger H_{u_2}) \; , \
        (H_{e_1}^\dagger H_{d_1}) (H_{e_1}^\dagger H_{e_2}) \; , \\ \nonumber
        &(H_{e_2}^\dagger H_{d_1}) (H_{e_2}^\dagger H_{e_3}) \; , \
        (H_{u_2} H_{d_3}) (H_{u_1}^\dagger H_{u_2}) \; , \\ \nonumber
        &(H_{d_3} H_{u_1}) (H_{d_1}^\dagger H_{d_3}) \; , \
        (H_{d_1} H_{u_1}) S^\dagger S^\dagger \; .
    \end{align}

%%%%%%%%%%%%%%%%%%%%%%%%%%%%%%%%%%%%%%%%%%%%%%%%%%%%%%%%
%%%%%%%%%%%%%%%%%%%%%%%%%%%%%%%%%%%%%%%%%%%%%%%%%%%%%%%%
\subsection{Comparison with KSVZ-Type Models}
\label{Subsec:Comparison_with_KSVZ-type_models}
\vs{-5mm}

\begin{figure*}
    \centering
    \includegraphics[width=\textwidth]{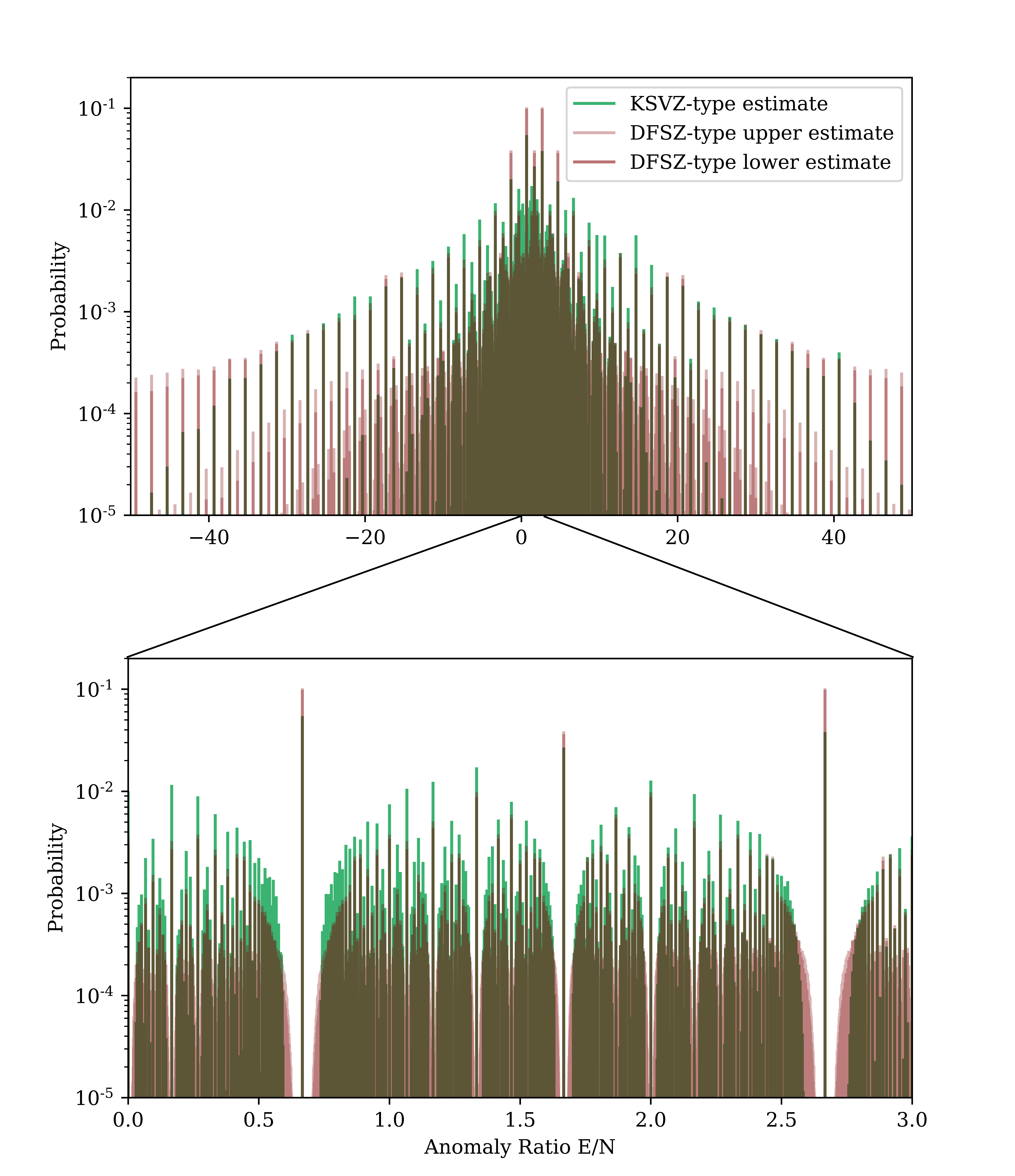}
    \caption{Comparison between anomaly ratio distributions for KSVZ-type and DFSZ-type axion models. The KSVZ-type estimate of \cite{Plakkot} includes all phenomenologically allowed models, adding and subtracting quark representations and assumes every model to be equally likely. Our DFSZ-type results include calculations for DFSZ$_2$ to DFSZ$_7$ and estimates for DFSZ$_8$ and DFSZ$_9$ giving equal probability to each $n_D$. For DFSZ-type, the different shades denote maximum and minimum for each bin under the two limits for DFSZ$_{8}$ and DFSZ$_9$ described above.}
    \label{fig:pdfcompare}
\end{figure*}

In \cite{Plakkot}, the authors add all anomaly ratios of phenomenologically allowed KSVZ-type models, irrespective of the number of quarks, allowing to add or subtract quark representations. This means that a single model with $N_Q = 9$ quarks, of which there are $> 1\times 10^5$, is deemed equally probable as a single model with $N_Q = 1$, of which there are only $15$. The distribution is therefore dominated by $7 \lesssim N_Q \lesssim 21$. If we used a similar approach for our DFSZ-type models, extrapolating the evolution of unique solutions with increasing $n_D$, the resulting distribution would be indistinguishable from the DFSZ$_9$ case. In Sec.~\ref{subsec:choices_for_a_statistical_interpretation}, we argued to instead use an approach in which all separate values for $n_D$ are equally probable. Since raw data was provided by \cite{Plakkot}, we are able to weight their KSVZ data in a way that gives equal probability to all values of $N_Q$.\footnote{So now the 15 models with $N_Q = 1$ combined are equally likely as all $> 1\times 10^5$ models with $N_Q = 9$ combined.}

Using this weighting, their data can be compared with our DFSZ results on a fair basis and we show the result in Fig.~\ref{fig:pdfcompare}. Nevertheless differences remain: The authors of \cite{Plakkot} were able to apply strict selection criteria, significantly reducing the number of viable models. We did not find similar stringent selection criteria, so our catalogue reflects the full set of models rather than a preferred set. A comparison between the two types of models or a combined axion band should therefore not be seen as final, but only as incorporating all selection criteria known so far. Also note that in our case, a model with higher $n_D$ is always less likely than a model with lower $n_D$, which can be seen as an appropriate penalty for introducing more degrees of freedom to the model. In our weighting scheme for KSVZ data from \cite{Plakkot} this is not the case, since for e.g. $N_Q=28$ they find only $510$ preferred models, much less than for $N_Q=9$, making a single model with $N_Q=28$ more likely than a single model with $N_Q=9$ in our approach.

One can clearly see the effect of the equal weights for all $n_D$ in Fig.~\ref{fig:pdfcompare} in the region around $E/N = 5/3$: The five $E/N$ values of the DFSZ$_2$ and DFSZ$_3$ models show highly elevated probability due to their big relative probabilities (compare Fig.~\ref{fig:pdfn3to9}). The effect of the two estimates for DFSZ$_8$ and DFSZ$_9$ only becomes substantial at low absolute probabilities and above $|E/N| \gtrsim 20$. We find the KSVZ results to also form a peaked structure similar to the DFSZ case, which only becomes visible in a very finely binned histogram. In fact, for $E/N$ values excluding DFSZ$_2$ and DFSZ$_3$, the DFSZ-type histograms are less peaked than the KSVZ-type ones, with decreased probability at moderately big $|E/N|$ and significantly increased probability for $|E/N| \gtrsim 40$. This trend does not translate to the biggest possible axion-photon coupling however. We find a maximal $|g_{a\gamma}|$ at $ \widehat{E/N} > -1216/3$, which is comparable to the KSVZ case for $N_Q \leq 9$ before any phenomenological constraints ($\widehat{E/N} = -1312/3$). 

Concerning models with smallest axion to photon couplings, in the following photophobic models are defined the same way as in \cite{Plakkot} by their anomaly ratio $E/N$ being compatible with vanishing $g_{a\gamma}$ within 1 sigma theoretical uncertainty (see Eq.~(\ref{Eq:AxionPhotonC})). Tab.~\ref{tab:catalogue} shows that there is no clear trend toward a higher or lower percentage of photophobic models with increasing $n_D$. As discussed in Sec.~\ref{Subsec:nD4-7} the anomaly ratio distribution becomes smoother with increasing $n_D$: Peaks become less pronounced and anomaly ratios with low probability become more likely. The absence of a clear trend hints at the photophobic region being right in the middle between those two extremes. Overall the percentage of photophobic models we find for DFSZ-type models with $n_D \leq 7$ is similar to the KSVZ case.

In both, KSVZ- and DFSZ-type of model probability distributions, the probability close to the highest peaks is strongly suppressed (Fig.~\ref{fig:pdfcompare}, bottom). The effect is less severe for DFSZ-type models than for KSVZ-type ones, because as noted before the former are less peaked if we subtract the effect of DFSZ$_2$ and DFSZ$_3$.

Upon closer inspection the distribution of KSVZ-type models is not symmetric around $5/3$ however, unlike the DFSZ-type one. Median and mean anomaly ratios are $E/N |_{\rm mean} = 1.43$ and $E/N |_{\rm median} = 1.30$ respectively, whereas for DFSZ-type models both are exactly $E/N |_{\rm mean} = E/N |_{\rm median} = 5/3$. These values remain unchanged, even if only considering the subset of $N_\DW =1$ models. The deviation from $5/3$ in the KSVZ-type models of \cite{Plakkot} may arise due to the phenomenological selection criteria they impose.

%%%%%%%%%%%%%%%%%%%%%%%%%%%%%%%%%%%%%%%%%%%%%%%%%%%%%%%%
%%%%%%%%%%%%%%%%%%%%%%%%%%%%%%%%%%%%%%%%%%%%%%%%%%%%%%%%
\section{IMPLICATIONS FOR AXION SEARCHES}
\label{sec:IMPACT_ON_AXION_SEARCHES}
\vs{-5mm}

\subsection[$C_{a\gamma}$ Cumulative Distribution Function]{$\mathbf{C_{a\gamma}}$ Cumulative Distribution Function}
\label{Subsec:gag_cdf}
\vs{-5mm}

We have so far derived probability mass functions for the anomaly ratio from theoretical assumptions for different DFSZ-type theories. To be able to understand the implications for axion searches, we need to map these $E/N$ distributions into $g_{a\gamma}$ space via Eq.~(\ref{Eq:AxionPhotonC}). In order to be independent of the axion mass we plot our results with respect to the unitless quantity $|\mathcal{C}_{a\gamma}|$ defined in Eq.~(\ref{Eq:AxionPhotonC}).

Traditionally two-sided axion bands centred around the region of maximal probability are given in this case \cite{Plakkot, DiLuzio:2017pfr, SLOAN201695, axionbands2, axionbands3}. However, usually an experiment is sensitive to all axion-photon couplings above a certain threshold $\left|\mathcal{C}_{a\gamma}\right|^{\rm min}$. We therefore deem it to be also relevant for experiments to post a one-sided limit that has to be reached in order to be sensitive to, e.g.~$68\%$ of all DFSZ-type models given a specific axion mass. For this purpose we use a cumulative distribution function (CDF) plotted against $\left|\mathcal{C}_{a\gamma}\right|$, which can be understood as the combined theoretical prior probability of models with $\left|\mathcal{C}_{a\gamma}\right|({\rm model}) > \left|\mathcal{C}_{a\gamma}\right|^{\rm min}$.

Since we are treating the anomaly ratio as a random variable coming from a distribution that we try to determine, we have to treat the second part of $\mathcal{C}_{a\gamma}$, the next-to-leading order QCD corrections $\mathcal{C}_{a\gamma \gamma}^{(0)}$, in the same way. We model its uncertainty as a normal distribution $\mathcal{N}(1.92, 0.04)$ with mean $1.92$ and standard deviation $0.04$. This smooths out steps in the CDF from high probability $E/N$ values, especially for anomaly ratios close to the mean value of $\mathcal{C}_{a\gamma \gamma}^{(0)}$.

%%%%%%%%%%%%%%%%%%%%%%%%%%%%%%%%%%%%%%%%%%%%%%%%%%%%%%%%
%%%%%%%%%%%%%%%%%%%%%%%%%%%%%%%%%%%%%%%%%%%%%%%%%%%%%%%%
\subsection{Experimental Constraints}
\label{Subsec:Experimental_Constraints}
\vs{-5mm}
%%%%%%%%%%%%%%%%%%%%%

Under the assumptions outlined above, we note that the anomaly ratios of the DFSZ$_2$ and DFSZ$_3$ models still are the most notable features in the probability distribution, even for all possible DFSZ models. However, since only one value of the anomaly ratio is realised in nature, reaching sensitivity to these models may be either not necessary or not sufficient.

\begin{figure}
    \centering
    \includegraphics[width=0.49\textwidth]{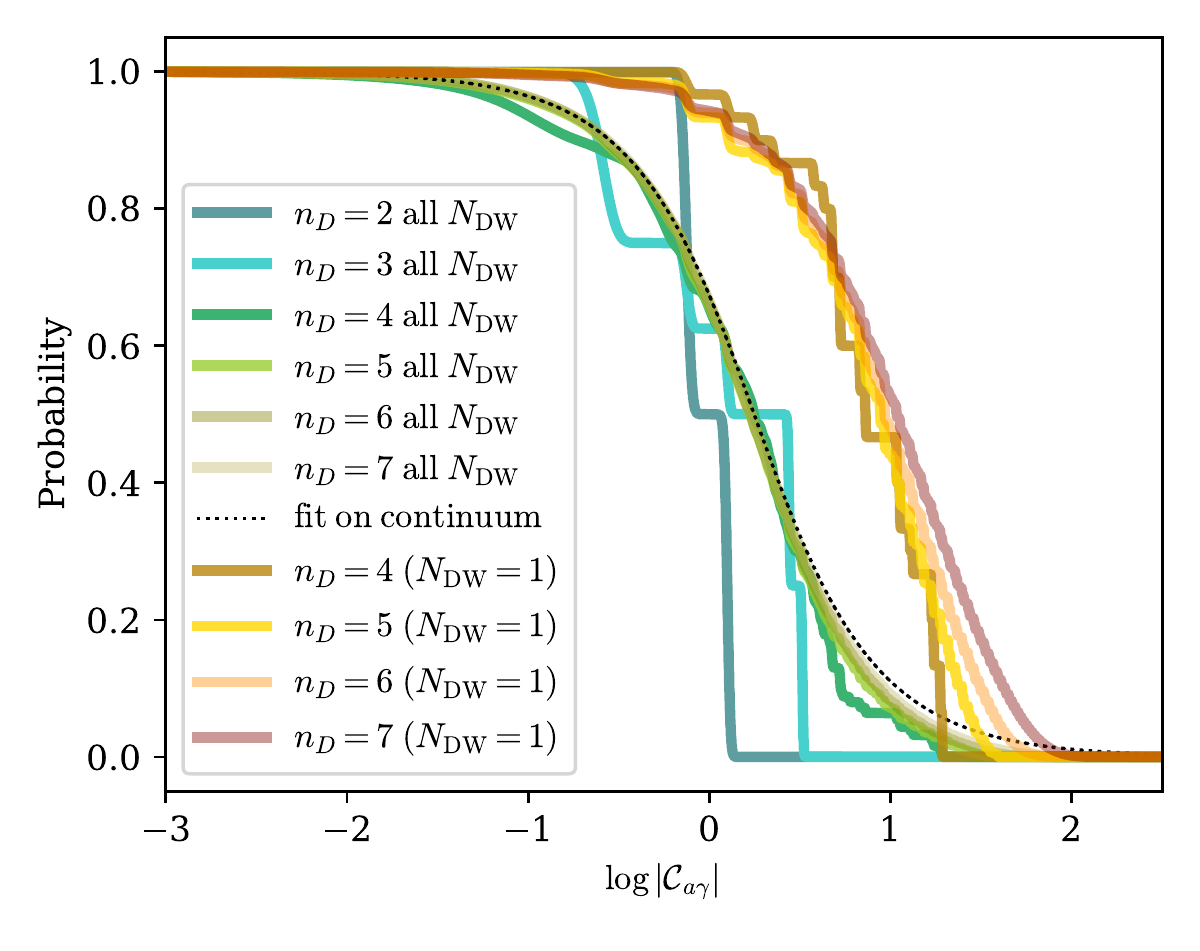}
    \caption{Cumulative probability of models with $|\mathcal{C}_{a\gamma}|$ higher than the indicated values. The plot includes DFSZ-type models of arbitrary domain wall number $N_\DW$ with DFSZ$_3$ to DFSZ$_7$ as well as $N_\DW = 1$ models for DFSZ$_4$ to DFSZ$_7$ (for smaller $n_D$ no $N_\DW = 1$ models exist). The CDFs become smoother with increasing $n_D$, with DFSZ$_6$ and DFSZ$_7$ already being almost indistinguishable. $N_\DW = 1$ models have systematically larger $|\mathcal{C}_{a\gamma}|$, shifted by almost one order of magnitude. The dashed line indicates the analytic fit on the continuum limit from Eq.~(\ref{Eq:cagfit}).}
    \label{fig:cdfDFSZ}
\end{figure}

Fig.~\ref{fig:cdfDFSZ} shows the resulting theoretical prior probability of DFSZ-type axion models with $|\mathcal{C}_{a\gamma}|$ higher than a specific value. We break the results down by possible values of $n_D$. Let us first discuss the ``all $N_\DW$"-case, in which the domain wall number does not present a meaningful selection criterion. DFSZ$_3$ models have zero probability above $\log |\mathcal{C}_{a\gamma}| \gtrsim 0.5$. Should an axion be found above this value that can be determined to be of DFSZ-type, this would imply the existence of $n_D > 3$ Higgs doublets. The impact of the prominent peaks of maximal probability between $E/N = -4/3$ or $E/N=14/3$ on the cumulative probability is only minor for theories with $n_D \geq 5$. Since the CDFs for DFSZ$_6$ and DFSZ$_7$ are already almost indistinguishable, we refrain from additionally plotting our estimates for higher $n_D$. In fact, the relative difference on $|\mathcal{C}_{a\gamma}|$ exclusion limits between our two ways of estimating the smoothness of the DFSZ$_{8}$ and DFSZ$_9$ distributions is below the percent level. For the purpose of $|\mathcal{C}_{a\gamma}|$ exclusion limits the two estimates are therefore virtually equivalent. In the following we use limit 2, the extrapolation estimate.

It is possible to obtain a reasonable analytic estimate for the cumulative probability distribution by going back to the analytic anomaly ratio fit from Eq.~(\ref{Eq:EoNfit}). For $|\mathcal{C}_{a\gamma}|$, it translates to
\begin{multline}
\label{Eq:cagfit}
    p\left( |\mathcal{C}_{a\gamma}| \right) = \\ 1 - \frac{\tan^{-1} \left[ \frac{4}{7} \left( |\mathcal{C}_{a\gamma}| - \frac{19}{75}\right) \right] + \tan^{-1} \left[ \frac{4}{7} \left( |\mathcal{C}_{a\gamma}| + \frac{19}{75}\right) \right]}{\pi},
\end{multline}
which is plotted as a dotted line in Fig.~\ref{fig:cdfDFSZ}.

Now contrast the full set of DFSZ$_4$ to DFSZ$_7$ models with the respective subsets having $N_\DW=1$. The latter models could be considered preferred in the post-inflationary scenario due to cosmological energy density arguments (see Sec.~\ref{Subsec:Selection_Criteria}). $N_\DW = 1$ models display $|\mathcal{C}_{a\gamma}|$ values almost an order of magnitude higher on average than the full set and are therefore much easier to detect. Similarly to the  ``all $N_{\DW}$"-case, higher $n_D$ values tend to have smoother distributions. It therefore seems reasonable to analogously introduce our two estimates where the difference with respect to the $|\mathcal{C}_{a\gamma}|$ limits between the two estimates is again below the percent level. We again use limit 2, the extrapolation estimate, in the following.

\begin{figure}
    \centering
    \includegraphics[width=0.48\textwidth]{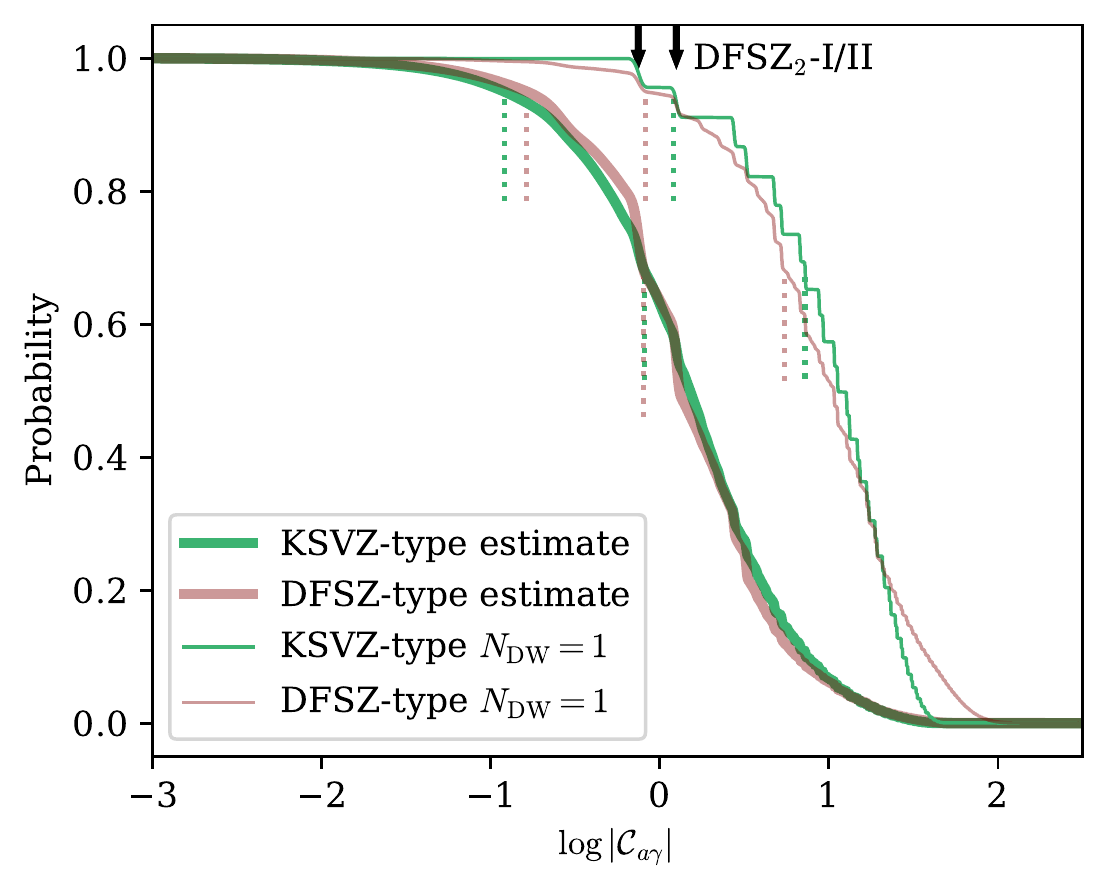}
    \caption{Cumulative probability of models with $|\mathcal{C}_{a\gamma}|$ higher than the indicated values for the complete set of DFSZ-type and KSVZ-type models as well as for models with $N_\DW = 1$ specifically (thin lines). One sided $95\%$ and $68\%$ limits for both cases are given with coloured vertical dotted lines. The arrows at the top indicate the location of DFSZ$_2$-I and DFSZ$_2$-II.}
    \label{fig:cdfcompare}
\end{figure}

\begin{figure}
    \centering
    \includegraphics[width=0.48\textwidth]{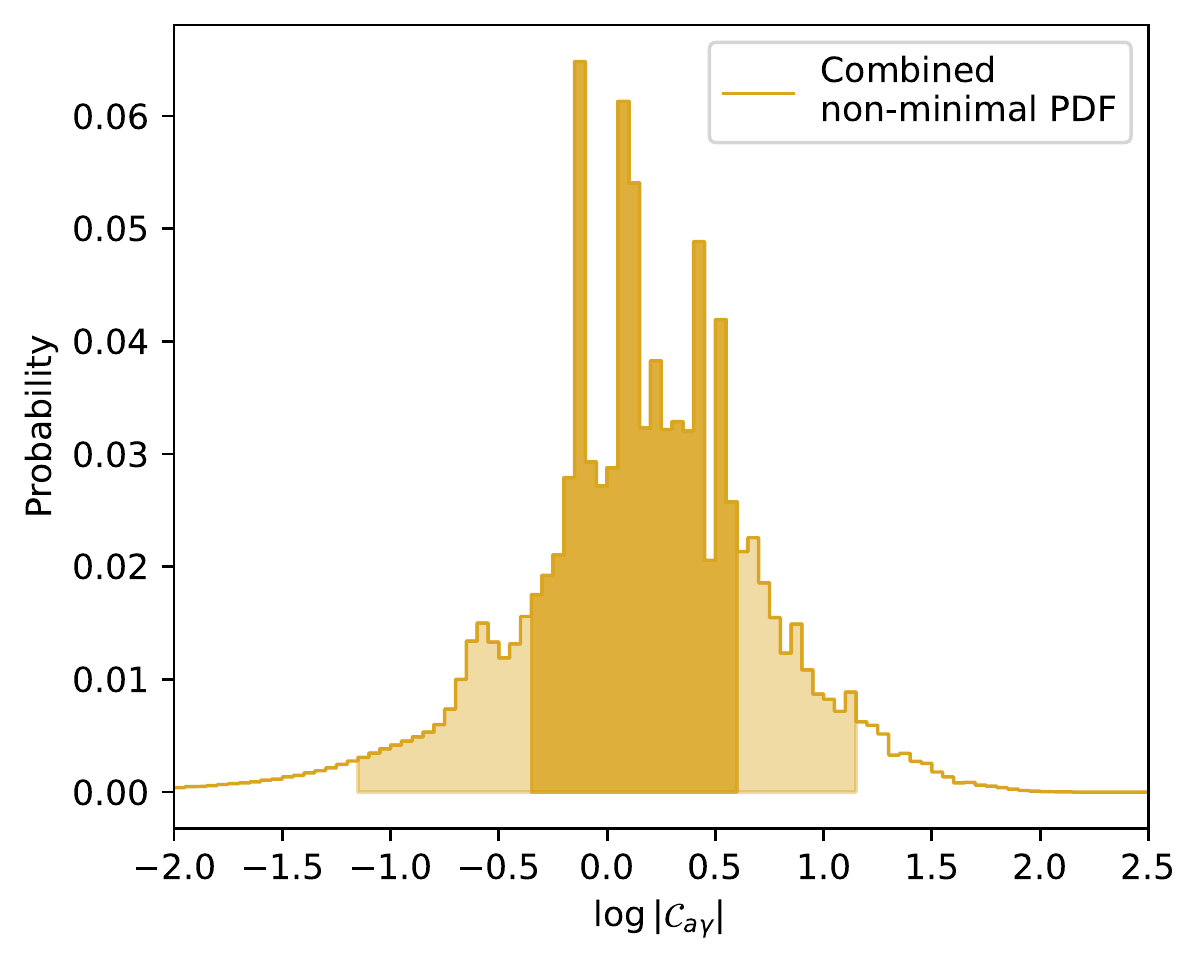}
    \caption{Probability density in $\log |\mathcal{C}_{a\gamma}|$-space of the combined DFSZ-type and KSVZ-type ``all $N_{\DW}$"-case. Central $68\%$ and $95\%$ regions used for Fig.~\ref{fig:gagexclusion} are indicated in different shades of yellow. Note that the underlying distribution is discrete and any illustration will in part depend on the binning chosen.}
    \label{fig:pdfcombined}
\end{figure}

We show Fig.~\ref{fig:cdfcompare} for a comparison of the CDFs for DFSZ- and KSVZ-type models. In general, both types are very similar for all $|\mathcal{C}_{a\gamma}|$. Only for DFSZ-type models with $N_\mathrm{DW}=1$ a significant percentage of models is above $\log |\mathcal{C}_{a\gamma}| \gtrsim 1.5$. The lines of $E/N = 2/3$ and $E/N = 8/3$ are clearly visible for DFSZ-type models, but also for KSVZ-type models. The relative difference between the $68\%$ limits of KSVZ\footnote{Again, considering all \textit{preferred} KSVZ-type models, see \cite{Plakkot} for more information.}- and DFSZ-type axions is only $\sim3\%$ and $\sim 19\%$ for the $95\%$ limits with the DFSZ limit being higher in the latter case. Taking into account possible effects from diverging theory assumptions, this relative difference can be seen as negligible. 

\begin{figure*}
    \centering
    \includegraphics[width=0.8\textwidth]{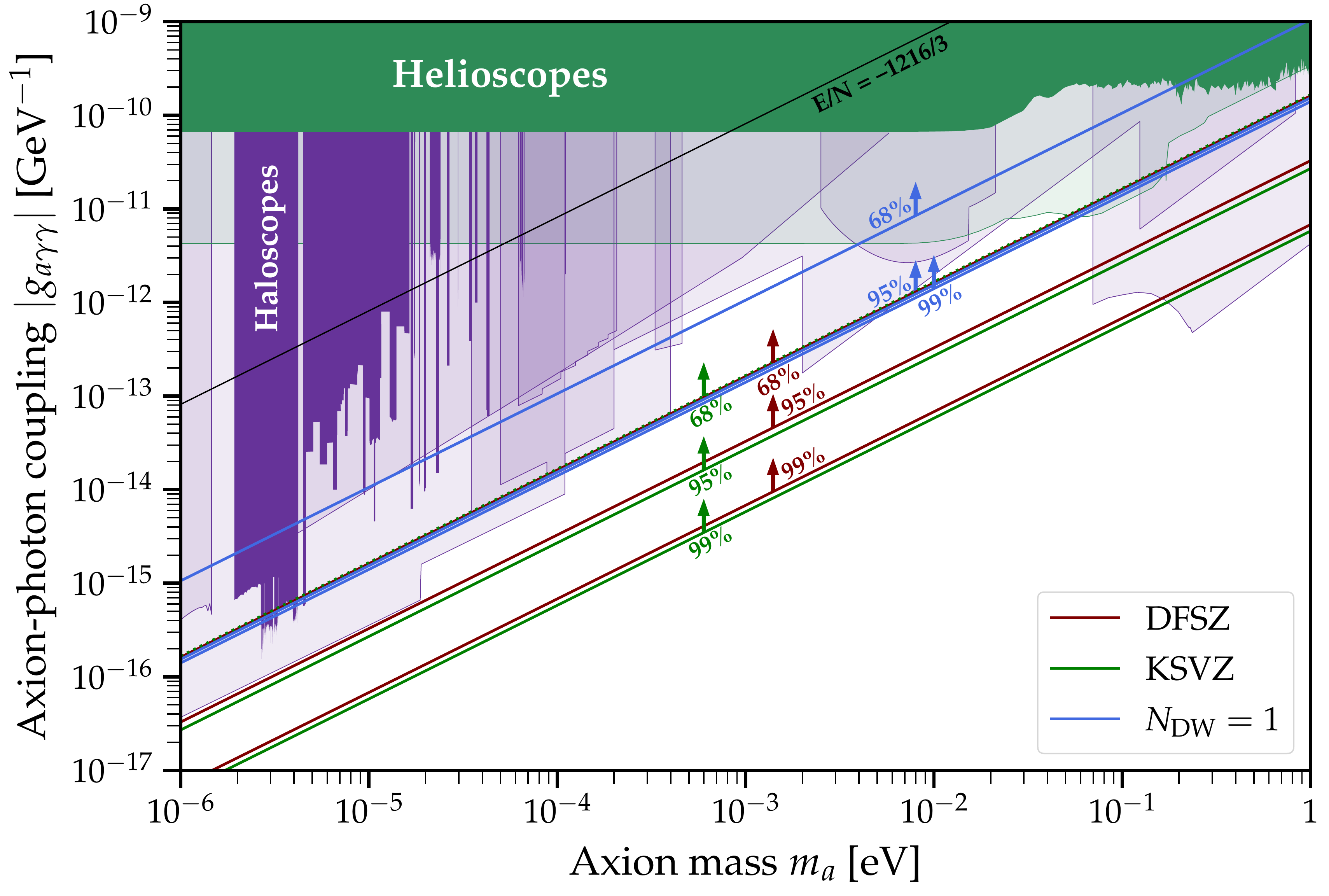}
    \includegraphics[width=0.8\textwidth]{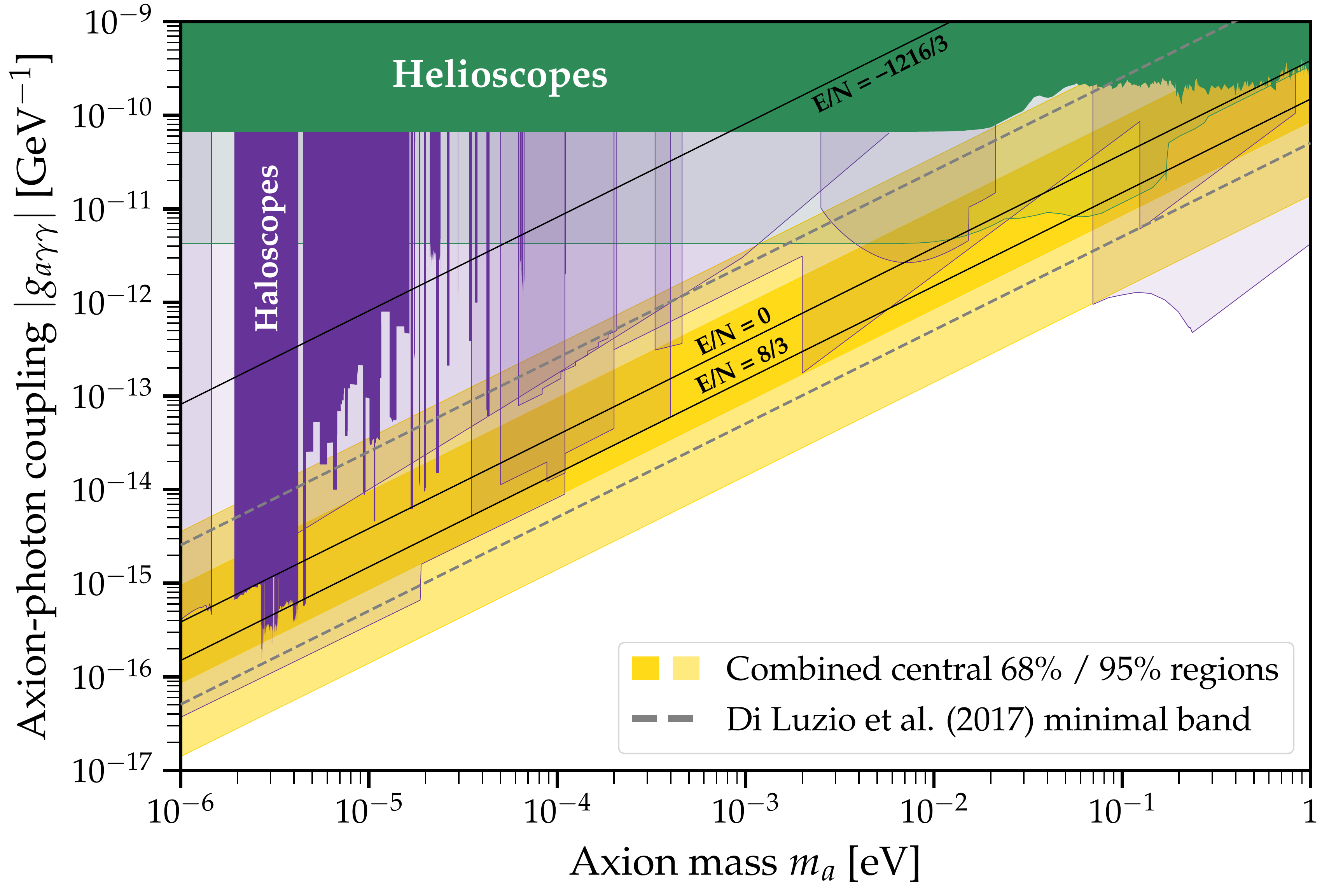}
    \caption{\textbf{Top:} $68\%$, $95\%$ and $99\%$ limits for the complete preferred KSVZ case \cite{Plakkot}, our complete DFSZ case (using extrapolation for DFSZ$_8$ and DFSZ$_9$) as well as the combined $N_\DW=1$ case. The highest DFSZ-type coupling found is shown in black ($E/N = -1216/3$). DFSZ$_2$-I and DFSZ$_2$-II roughly coincide with the $68\%$ limit of the complete DFSZ case and the $95\%$ limit of the $N_\DW = 1$ case, respectively. \textbf{Bottom:} Central $68\%$ and $95\%$ regions for the case combining all preferred KSVZ and all DFSZ models together with a previous band from Di Luzio \textit{et al}. \cite{DiLuzio:2017pfr} for comparison. We show helioscope limits and forecasts \cite{Shilon:2012te, CAST:2017uph, 2019JCAP...06..047A} in green as well as limits and forecasts from various haloscope experiments \Haloscopes \ in purple. For reference we also show the $E/N=0$ and $E/N=8/3$ lines in black. All experimental limits shown here are Frequentist in nature and should therefore only be seen as a rough comparison with respect to our Bayesian prior results. For the full cumulative probability from which the three limits shown in the top panel are taken, see Fig.~\ref{fig:cdfcompare}, and for the combined probability density from which the band in the bottom panel is derived, see Fig.~\ref{fig:pdfcombined}. (Plotted with tools by O'Hare \cite{AxionLimits}.)}
    \label{fig:gagexclusion}
\end{figure*}

While the investigation of different theoretical assumptions is beyond the scope of this paper, note that other assumptions on the full set of models only modify the relative importance of the prominent DFSZ$_2$ and DFSZ$_3$ peaks. Consider for example a different definition of multiplicity based on minimal potentials. This dramatically increases their probability mass but does not shift the overall cumulative probability to higher or lower $|\mathcal{C}_{a\gamma}|$ values. In this sense any variation of theoretical assumptions (excluding model selection criteria) should lie between the cumulative probabilities of the $n_D=2$ and continuum cases.

Translating these limits to $g_{a\gamma}$ over a range of axion masses, we obtain Fig.~\ref{fig:gagexclusion} (top). An experimental exclusion limit touching the $68\%$ line excludes $68\%$ of the probability mass over the model space under the assumptions outlined above given a specific mass range. An experiment targeting sensitivity down to the $95\%$ line will be sensitive to $95\%$ of the probability for all models in the targeted mass range. We include these and the $99\%$ limit for DFSZ-type as well as KSVZ-type models and the combined case of $N_\DW = 1$. In black we also include the maximal $\widehat{E/N}$ value we found for DFSZ$_9$. In addition to being excluded by experiments for a large fraction of the $m_a$ range, this model may likely also be subject to phenomenological constraints (see Sec.~\ref{Subsec:Selection_Criteria}).

\begin{table}
  \caption{$|\mathcal{C}_{a\gamma}|$ lower prior limits for selected combinations of models. All limits shown are one sided, so a central $68\%$ band can be constructed with values given for $16\%$ and $84\%$ and similar for $95\%$. \textbf{KSVZ} denotes re-weighted results from \cite{Plakkot}, \textbf{DFSZ} results from this paper. Both are combined with equal probability for the case \textbf{Combined}. The combination only considering models with DW number of unity is shown as $\mathbf{N_\mathrm{DW}=1}$. }
\begin{tabular}{l | r| r? r| r? r| r} 
 \toprule
 \multicolumn{3}{l}{$|\mathcal{C}_{a\gamma}|$} & \multicolumn{2}{?c?}{$68\%$ band} & \multicolumn{2}{c}{$95\%$ band} \\
 \midrule
 One-sided limit & $68\%$ & $95\%$ & $16\%$ &$84\%$ & $2.5\%$ & $97.5\%$ \\ \midrule
KSVZ               & 0.833  &  0.135 &   4.684  &  0.427 &  15.274 &   0.068 \\
DFSZ               & 0.809  &  0.164 &   4.529  &  0.482 &  19.272 &    0.08 \\
Combined           & 0.819  &  0.148 &   4.593  &  0.451 &  17.285 &   0.074 \\
$N_\mathrm{DW}=1$  & 5.294  &  0.769 &  22.773  &  2.733 &  36.729 &   0.731 \\ \bottomrule
\end{tabular}
  \label{tab:limits}
\end{table}

With this work it is now possible for the first time to give values of one-sided limits or axion bands for the combined KSVZ and DFSZ case, assuming a DFSZ-type axion to be equally likely as a KSVZ-type one. The associated PDF for the combined ``all $N_{\DW}$"-case is shown in Fig.~\ref{fig:pdfcombined}. In $\log |\mathcal{C}_{a\gamma}|$-space with the relatively course binning chosen, the distributions look roughly Gaussian with the exception of several notable peaks, at $E/N = 5/3, 8/3, 2/3, 14/3$ and $-4/3$ (from left to right). Note, however, that the true underlying distribution is comprised out of a multitude of delta peaks and thus fundamentally discrete. Central $68\%$ and $95\%$ bands from this distribution are used in Fig.~\ref{fig:gagexclusion} (bottom) together with a previous estimate for the same band from \cite{DiLuzio:2017pfr}. Previous work was either limited to very few extensions of DFSZ-type \cite{DiLuzio:2017pfr} or the KSVZ case \cite{Plakkot}. Even now many caveats have to be kept in mind, like the imprecise prediction for DFSZ$_8$ and DFSZ$_9$ models or the lack of selection criteria in the DFSZ case. Acknowledging this, we nevertheless deem it useful to provide usable data of typical limits and bands for a variety of scenarios. An overview can be found in Tab.~\ref{tab:limits} and more detailed information is hosted on the website ``zenodo" together with the model catalogues (see end of Sec.~\ref{sec:Summary_and_Outlook} for links).

%%%%%%%%%%%%%%%%%%%%%%%%%%%%%%%%%%%%%%%%%%%%%%%%%%%%%%%%
%%%%%%%%%%%%%%%%%%%%%%%%%%%%%%%%%%%%%%%%%%%%%%%%%%%%%%%%
\section{SUMMARY AND OUTLOOK}
\label{sec:Summary_and_Outlook}
\vs{-5mm}

The PQ mechanism is the most commonly considered solution to the strong CP problem and the appearing Goldstone boson, the axion, is one of the most promising dark matter candidates. While the axion solves the strong CP problem in a model independent way, its low-energy couplings depend on UV-models. With the booming axion experimental program, an identification of all these models within the two large classes of invisible axion models and the extraction of predictions for experiments are of high importance. In this work, we have systematically calculated the axion-photon coupling for a large number of DFSZ-type models. We give limits that have to be reached in order to be sensitive to a certain fraction of the probability mass of these models.

We have started by discussing (phenomenological) selection criteria, such as the absence of FCNCs and the DW problem, to extract preferred DFSZ-type models. In contrast to the KSVZ axions, where all selection criteria follow from cosmological bounds on additional fermions, for DFSZ-type axions we have not find criteria with a sufficient level of generality, merely desirable features.

Next, we have put forth a recipe for calculating all anomaly ratios and hence all axion-photon couplings. This recipe is based on the fact that the PQ charges are not free but fixed by linear consistency and phenomenology relations. For the sake of calculating the anomaly ratio, this reduces the procedure of DFSZ-type model building to solving LSEs. Thus, systematically going through all Yukawa sectors and solving all possible LSEs for each, we have derived all possible anomaly ratios for up to seven Higgs doublets.

In addition, by counting how many models lead to a certain anomaly ratio and establishing relative probabilities of these models, we have been able to assign probabilities to each anomaly ratio. For this counting of models we have considered as a model the Lagrangian that arises by combining different potentials that give rise to the same set of PQ charges and by adding the Yukawa couplings compatible with the resulting set of PQ charges. In this way we have taken into account the general mantra that all terms allowed by symmetry should be included and avoid overcounting.

The resulting anomaly ratio distributions have their median at $E/N=5/3$, their maximum values at $E/N=2/3$ and $E/N=8/3$, and a characteristic shape that is similar to the one of KSVZ-type models. We have explained these observations by thinking of the resulting sets of PQ charges as discrete charge distributions with uniform probability and symmetry around zero.

While in principle our recipe works for an arbitrary number of Higgs doublets, the necessary computational time becomes too large for eight or more doublets. Simple sampling of potential terms leads to a significant bias, so that we have constructed estimates for $n_D > 7$ based on the expected smoothness of the distributions. Moreover, by using an incremental construction procedure, we have been able to find a maximal anomaly ratio that is more than a factor of two higher than in previous estimates \cite{DiLuzio:2017pfr}.

Regarding the axion experimental program the anomaly ratio distributions confirm the experimental importance of the values dictated by the minimal DFSZ models, namely $E/N = 2/3$ and $E/N=8/3$, since they are also favoured for every number of Higgs doublets (except $n_D=3$ with the Weinberg-Glashow-Paschos condition imposed). However, it also shows that plenty of viable parameter space lies above and below these lines. Overall, this means that a non-observation at these favoured values is not enough to declare the axion excluded, while an observation above these values would be a hint for more than one additional Higgs doublet from the DFSZ-type point of view. The statistical interpretation also reveals that both, KSVZ and DFSZ models, set very similar sensitivity requirements on experiments.

For $n_D \geq 4$ we have found a subset of models with DW number $N_\DW = 1$, making DFSZ-type models theoretically more viable in post-inflationary scenarios. This subset even displays a significantly enhanced axion-photon coupling compared to the minimal scenarios for both invisible axion classes, hence making these models on average easier to probe.

Our analysis can be extended in multiple directions. For instance, it would be interesting to perform a similar analysis for models with a right-handed neutrino or KSVZ/DFSZ hybrid models, namely models with additional Higgs singlets, Higgs doublets, and heavy quarks. From our arguments regarding the anomaly ratio from a mathematical point of view, even though Eq.~(\ref{Eq:EoverN}) would change, we expect a similar shape of the resulting distributions and axion mass vs. axion-photon coupling exclusion lines. This expectation does, however, not make an explicit analysis dispensable. Furthermore, it would be interesting to investigate other axion couplings, such as the axion-electron coupling. While the other couplings depend on the VEVs of the Higgs doublets, which makes the parameter space higher dimensional, the perturbative range of the top and bottom Yukawa couplings \cite{Bjorkeroth:2019jtx} or phenomenological constraints could be used to give reasonable limits. Additionally, it would be desirable to find a better estimate for the anomaly distribution of eight or more doublets, or even an unbiased way to calculate it.

Finally, it is interesting to mention that with the identification and classification of both large classes of invisible axion models, also a comparison with other classes of axion models is possible. For instance there exists the two-form implementation of the QCD axion \cite{Dvali:2005an, PhysRevD.105.085020, Dvali:2022fdv}. This intrinsically is a gauge formulation of the axion and as such no explicit PQ violating processes are possible. This is not true for the ordinary invisible axion. Hence, should the axion be detected, PQ violating processes represent an interesting feature to not only distinguish the two-form axion from the ordinary invisible axion but to completely eliminate it.

Lastly, it should be said that our analysis is useful for axion searches irrespective of the statistical interpretation. By providing all possible $E/N$ values for up to seven doublets and a full catalogue for up to five doublets, in the case of a detection one can proceed to do hypothesis testing with the compatible models. Since all $E/N$ values for preferred KSVZ models are also known, this could be used for the purpose of model comparison between these two model classes. 

Hence, with or without a statistical perspective, our work presents another step forward in the understanding and mapping of the landscape of axion models.

Our generating code can be found at \url{https://github.com/jhbdiehl/DFSZforest}, the model catalogues and axion limits/ bands together with usable Bayesian theory priors at \url{https://doi.org/10.5281/zenodo.7656939}.

%%%%%%%%%%%%%%%%%%%%%%%%%%%%%%%%%%%%%%%%%%%%%%%%%%%%%%%%
%%%%%%%%%%%%%%%%%%%%%%%%%%%%%%%%%%%%%%%%%%%%%%%%%%%%%%%%
\acknowledgments
\vs{-5mm}

Special thanks goes to Oliver Schulz for crucial input on the Julia implementation of our LSEs. Without him our code would still be running! We are grateful to him, Gia Dvali, Béla Majorovits, Georg Raffelt, and Frank Steffen for very useful discussions and helpful comments on the manuscript. We thank Vaisakh Plakkot and Sebastian Hoof for discussions which started this project and their excellent groundwork on KSVZ models. This project was initiated at the 2021 DPG Bad Honnef WISP summer school, and we therefore thank the organisers for their hospitality.

\newpage

%%%%%%%%%%%%%%%%%%%%%%%%%%%%%%%%%%%%%%%%%%%%%%%%%%%%%%%%%%%
%%%%%%%%%%%%%%%%%%%%%%%%%%%%%%%%%%%%%%%%%%%%%%%%%%%%%%%%
\setlength{\bibsep}{5pt}
\setstretch{1}
\bibliographystyle{utphys86}
\bibliography{main}

\providecommand{\href}[2]{#2}\begingroup \begin{thebibliography}{10}

\bibitem{nEDM2015}
J.~Pendlebury, S.~Afach, N.~Ayres, {\em et~al.}, ``Revised experimental upper
  limit on the electric dipole moment of the neutron,''
  \href{http://dx.doi.org/10.1103/physrevd.92.092003}{{\em Physical Review D}
  {\bfseries 92} no.~9, (Nov, 2015) }.
  \url{http://dx.doi.org/10.1103/PhysRevD.92.092003}.

\bibitem{Weinberg:1996kr}
S.~Weinberg, {\em {The quantum theory of fields. Vol. 2: Modern applications}}.
\newblock 1996.
\newblock \url{http://www.slac.stanford.edu/spires/find/hep/www?key=3763846}.
\newblock Cambridge, UK: Univ. Pr. (1996) 489 p.

\bibitem{ThetaCorrections}
J.~Ellis and M.~K. Gaillard, ``Strong and weak cp violation,''
  \href{http://dx.doi.org/https://doi.org/10.1016/0550-3213(79)90297-9}{{\em
  Nuclear Physics B} {\bfseries 150} (1979) 141--162}.
  \url{https://www.sciencedirect.com/science/article/pii/0550321379902979}.

\bibitem{Dvali:2013eja}
G.~Dvali and C.~Gomez, ``{Quantum Compositeness of Gravity: Black Holes, AdS
  and Inflation},'' \href{http://dx.doi.org/10.1088/1475-7516/2014/01/023}{{\em
  JCAP} {\bfseries 01} (2014) 023},
  \href{http://arxiv.org/abs/1312.4795}{{\ttfamily arXiv:1312.4795 [hep-th]}}.

\bibitem{Dvali:2018jhn}
G.~Dvali, C.~Gomez, and S.~Zell, ``{Quantum Breaking Bound on de Sitter and
  Swampland},'' \href{http://dx.doi.org/10.1002/prop.201800094}{{\em Fortsch.
  Phys.} {\bfseries 67} no.~1-2, (2019) 1800094},
  \href{http://arxiv.org/abs/1810.11002}{{\ttfamily arXiv:1810.11002
  [hep-th]}}.

\bibitem{Dvali:2020etd}
G.~Dvali, ``{$S$-Matrix and Anomaly of de Sitter},''
  \href{http://dx.doi.org/10.3390/sym13010003}{{\em Symmetry} {\bfseries 13}
  no.~1, (2020) 3}, \href{http://arxiv.org/abs/2012.02133}{{\ttfamily
  arXiv:2012.02133 [hep-th]}}.

\bibitem{Dvali:2018dce}
G.~Dvali, C.~Gomez, and S.~Zell, ``{A Proof of the Axion?},''
  \href{http://arxiv.org/abs/1811.03079}{{\ttfamily arXiv:1811.03079
  [hep-th]}}.

\bibitem{Dvali:2022fdv}
G.~Dvali, ``{Strong-$CP$ with and without gravity},''
  \href{http://arxiv.org/abs/2209.14219}{{\ttfamily arXiv:2209.14219
  [hep-ph]}}.

\bibitem{PQMechanism}
R.~D. Peccei and H.~R. Quinn, ``$\mathrm{CP}$ conservation in the presence of
  pseudoparticles,'' \href{http://dx.doi.org/10.1103/PhysRevLett.38.1440}{{\em
  Phys. Rev. Lett.} {\bfseries 38} (Jun, 1977) 1440--1443}.
  \url{https://link.aps.org/doi/10.1103/PhysRevLett.38.1440}.

\bibitem{PQMechanism2}
R.~D. Peccei and H.~R. Quinn, ``Constraints imposed by $\mathrm{CP}$
  conservation in the presence of pseudoparticles,''
  \href{http://dx.doi.org/10.1103/PhysRevD.16.1791}{{\em Phys. Rev. D}
  {\bfseries 16} (Sep, 1977) 1791--1797}.
  \url{https://link.aps.org/doi/10.1103/PhysRevD.16.1791}.

\bibitem{WeinbergAxion}
S.~Weinberg, ``A new light boson?''
  \href{http://dx.doi.org/10.1103/PhysRevLett.40.223}{{\em Phys. Rev. Lett.}
  {\bfseries 40} (Jan, 1978) 223--226}.
  \url{https://link.aps.org/doi/10.1103/PhysRevLett.40.223}.

\bibitem{Wilczek:1977pj}
F.~Wilczek, ``{Problem of Strong $P$ and $T$ Invariance in the Presence of
  Instantons},''
\href{http://dx.doi.org/10.1103/PhysRevLett.40.279}{{\em Phys. Rev. Lett.}
  {\bfseries 40} (1978) 279--282}.
%%CITATION = PRLTA,40,279;%%.

\bibitem{SREDNICKI1985689}
M.~Srednicki, ``Axion couplings to matter: (i). cp-conserving parts,''
  \href{http://dx.doi.org/https://doi.org/10.1016/0550-3213(85)90054-9}{{\em
  Nuclear Physics B} {\bfseries 260} no.~3, (1985) 689--700}.
  \url{https://www.sciencedirect.com/science/article/pii/0550321385900549}.

\bibitem{DFSZ1}
A.~R. Zhitnitsky, ``{On Possible Suppression of the Axion Hadron Interactions.
  (In Russian)},'' {\em Sov. J. Nucl. Phys.} {\bfseries 31} (1980) 260.

\bibitem{DFSZ2}
M.~Dine, W.~Fischler, and M.~Srednicki, ``A simple solution to the strong cp
  problem with a harmless axion,''
  \href{http://dx.doi.org/https://doi.org/10.1016/0370-2693(81)90590-6}{{\em
  Physics Letters B} {\bfseries 104} no.~3, (1981) 199--202}.
  \url{https://www.sciencedirect.com/science/article/pii/0370269381905906}.

\bibitem{KSVZ1}
J.~E. Kim, ``{Weak Interaction Singlet and Strong CP Invariance},''
  \href{http://dx.doi.org/10.1103/PhysRevLett.43.103}{{\em Phys. Rev. Lett.}
  {\bfseries 43} (1979) 103}.

\bibitem{KSVZ2}
M.~A. Shifman, A.~I. Vainshtein, and V.~I. Zakharov, ``{Can Confinement Ensure
  Natural CP Invariance of Strong Interactions?},''
  \href{http://dx.doi.org/10.1016/0550-3213(80)90209-6}{{\em Nucl. Phys. B}
  {\bfseries 166} (1980) 493--506}.

\bibitem{DiLuzioAxionLandscape}
L.~Di~Luzio, M.~Giannotti, E.~Nardi, and L.~Visinelli, ``The landscape of qcd
  axion models,'' \href{http://dx.doi.org/10.1016/j.physrep.2020.06.002}{{\em
  Physics Reports} {\bfseries 870} (Jul, 2020) 1–117}.
  \url{http://dx.doi.org/10.1016/j.physrep.2020.06.002}.

\bibitem{DiLuzio:2017pfr}
L.~Di~Luzio, F.~Mescia, and E.~Nardi, ``{Window for preferred axion models},''
  \href{http://dx.doi.org/10.1103/PhysRevD.96.075003}{{\em Phys. Rev. D}
  {\bfseries 96} no.~7, (2017) 075003},
  \href{http://arxiv.org/abs/1705.05370}{{\ttfamily arXiv:1705.05370
  [hep-ph]}}.

\bibitem{Plakkot}
V.~{Plakkot} and S.~{Hoof}, ``{Anomaly ratio distributions of hadronic axion
  models with multiple heavy quarks},''
  \href{http://dx.doi.org/10.1103/PhysRevD.104.075017}{{\em \prd} {\bfseries
  104} no.~7, (Oct., 2021) 075017},
  \href{http://arxiv.org/abs/2107.12378}{{\ttfamily arXiv:2107.12378
  [hep-ph]}}.

\bibitem{GrillidiCortona:2015jxo}
G.~Grilli~di Cortona, E.~Hardy, J.~Pardo~Vega, and G.~Villadoro, ``{The QCD
  axion, precisely},'' \href{http://dx.doi.org/10.1007/JHEP01(2016)034}{{\em
  JHEP} {\bfseries 01} (2016) 034},
  \href{http://arxiv.org/abs/1511.02867}{{\ttfamily arXiv:1511.02867
  [hep-ph]}}.

\bibitem{THOOFT1986357}
G.~{'t Hooft}, ``How instantons solve the u(1) problem,''
  \href{http://dx.doi.org/https://doi.org/10.1016/0370-1573(86)90117-1}{{\em
  Physics Reports} {\bfseries 142} no.~6, (1986) 357--387}.
  \url{https://www.sciencedirect.com/science/article/pii/0370157386901171}.

\bibitem{PhysRevD.17.2717}
C.~G. Callan, R.~Dashen, and D.~J. Gross, ``Toward a theory of the strong
  interactions,'' \href{http://dx.doi.org/10.1103/PhysRevD.17.2717}{{\em Phys.
  Rev. D} {\bfseries 17} (May, 1978) 2717--2763}.
  \url{https://link.aps.org/doi/10.1103/PhysRevD.17.2717}.

\bibitem{Ernst:2018bib}
A.~Ernst, A.~Ringwald, and C.~Tamarit, ``{Axion Predictions in $SO(10)\times
  U(1)_{\rm PQ}$ Models},''
  \href{http://dx.doi.org/10.1007/JHEP02(2018)103}{{\em JHEP} {\bfseries 02}
  (2018) 103}, \href{http://arxiv.org/abs/1801.04906}{{\ttfamily
  arXiv:1801.04906 [hep-ph]}}.

\bibitem{QCDINstantonsFiniteTemp}
D.~J. Gross, R.~D. Pisarski, and L.~G. Yaffe, ``Qcd and instantons at finite
  temperature,'' \href{http://dx.doi.org/10.1103/RevModPhys.53.43}{{\em Rev.
  Mod. Phys.} {\bfseries 53} (Jan, 1981) 43--80}.
  \url{https://link.aps.org/doi/10.1103/RevModPhys.53.43}.

\bibitem{DWProblem}
P.~Sikivie, ``Axions, domain walls, and the early universe,''
  \href{http://dx.doi.org/10.1103/PhysRevLett.48.1156}{{\em Phys. Rev. Lett.}
  {\bfseries 48} (Apr, 1982) 1156--1159}.
  \url{https://link.aps.org/doi/10.1103/PhysRevLett.48.1156}.

\bibitem{Vilenkin:2000jqa}
A.~Vilenkin and E.~P.~S. Shellard, {\em {Cosmic Strings and Other Topological
  Defects}}.
\newblock Cambridge University Press, 7, 2000.

\bibitem{Dvali:1995cc}
G.~R. Dvali and G.~Senjanovic, ``{Is there a domain wall problem?},''
  \href{http://dx.doi.org/10.1103/PhysRevLett.74.5178}{{\em Phys. Rev. Lett.}
  {\bfseries 74} (1995) 5178--5181},
  \href{http://arxiv.org/abs/hep-ph/9501387}{{\ttfamily arXiv:hep-ph/9501387}}.

\bibitem{LAZARIDES198221}
G.~Lazarides and Q.~Shafi, ``Axion models with no domain wall problem,''
  \href{http://dx.doi.org/https://doi.org/10.1016/0370-2693(82)90506-8}{{\em
  Physics Letters B} {\bfseries 115} no.~1, (1982) 21--25}.
  \url{https://www.sciencedirect.com/science/article/pii/0370269382905068}.

\bibitem{Pich:2011nh}
A.~Pich, \href{http://dx.doi.org/10.5170/CERN-2013-003.119}{``{Flavour Physics
  and CP Violation},''} in {\em {6th CERN-Latin-American School of High-Energy
  Physics}}, pp.~119--144.
\newblock 2013.
\newblock \href{http://arxiv.org/abs/1112.4094}{{\ttfamily arXiv:1112.4094
  [hep-ph]}}.

\bibitem{Ivanov:2017dad}
I.~P. Ivanov, ``{Building and testing models with extended Higgs sectors},''
  \href{http://dx.doi.org/10.1016/j.ppnp.2017.03.001}{{\em Prog. Part. Nucl.
  Phys.} {\bfseries 95} (2017) 160--208},
  \href{http://arxiv.org/abs/1702.03776}{{\ttfamily arXiv:1702.03776
  [hep-ph]}}.

\bibitem{PhysRevD.15.1958}
S.~L. Glashow and S.~Weinberg, ``Natural conservation laws for neutral
  currents,'' \href{http://dx.doi.org/10.1103/PhysRevD.15.1958}{{\em Phys. Rev.
  D} {\bfseries 15} (Apr, 1977) 1958--1965}.
  \url{https://link.aps.org/doi/10.1103/PhysRevD.15.1958}.

\bibitem{PhysRevD.15.1966}
E.~A. Paschos, ``Diagonal neutral currents,''
  \href{http://dx.doi.org/10.1103/PhysRevD.15.1966}{{\em Phys. Rev. D}
  {\bfseries 15} (Apr, 1977) 1966--1972}.
  \url{https://link.aps.org/doi/10.1103/PhysRevD.15.1966}.

\bibitem{Gogberashvili:1991ws}
M.~Y. Gogberashvili and G.~R. Dvali, ``{Hierarchy at Yukawa constants and K0
  anti-K0, B0 anti-B0 oscillations in the model with two Higgs doublets. (In
  Russian)},'' {\em Sov. J. Nucl. Phys.} {\bfseries 53} (1991) 491--492.

\bibitem{Pich:2009sp}
A.~Pich and P.~Tuzon, ``{Yukawa Alignment in the Two-Higgs-Doublet Model},''
  \href{http://dx.doi.org/10.1103/PhysRevD.80.091702}{{\em Phys. Rev. D}
  {\bfseries 80} (2009) 091702},
  \href{http://arxiv.org/abs/0908.1554}{{\ttfamily arXiv:0908.1554 [hep-ph]}}.

\bibitem{deMedeirosVarzielas:2019dyu}
I.~de~Medeiros~Varzielas and J.~Talbert, ``{FCNC-free multi-Higgs-doublet
  models from broken family symmetries},''
  \href{http://dx.doi.org/10.1016/j.physletb.2019.135091}{{\em Phys. Lett. B}
  {\bfseries 800} (2020) 135091},
  \href{http://arxiv.org/abs/1908.10979}{{\ttfamily arXiv:1908.10979
  [hep-ph]}}.

\bibitem{PhysRevD.35.3484}
T.~P. Cheng and M.~Sher, ``Mass-matrix ansatz and flavor nonconservation in
  models with multiple higgs doublets,''
  \href{http://dx.doi.org/10.1103/PhysRevD.35.3484}{{\em Phys. Rev. D}
  {\bfseries 35} (Jun, 1987) 3484--3491}.
  \url{https://link.aps.org/doi/10.1103/PhysRevD.35.3484}.

\bibitem{PhysRevD.103.075026}
S.~Carrolo, J.~C. Rom\~ao, J.~a.~P. Silva, and F.~Vaz\~ao, ``Symmetry and
  decoupling in multi-higgs boson models,''
  \href{http://dx.doi.org/10.1103/PhysRevD.103.075026}{{\em Phys. Rev. D}
  {\bfseries 103} (Apr, 2021) 075026}.
  \url{https://link.aps.org/doi/10.1103/PhysRevD.103.075026}.

\bibitem{Penuelas:2017ikk}
A.~Pe\~nuelas and A.~Pich, ``{Flavour alignment in multi-Higgs-doublet
  models},'' \href{http://dx.doi.org/10.1007/JHEP12(2017)084}{{\em JHEP}
  {\bfseries 12} (2017) 084}, \href{http://arxiv.org/abs/1710.02040}{{\ttfamily
  arXiv:1710.02040 [hep-ph]}}.

\bibitem{Farina:2016tgd}
M.~Farina, D.~Pappadopulo, F.~Rompineve, and A.~Tesi, ``{The photo-philic QCD
  axion},'' \href{http://dx.doi.org/10.1007/JHEP01(2017)095}{{\em JHEP}
  {\bfseries 01} (2017) 095}, \href{http://arxiv.org/abs/1611.09855}{{\ttfamily
  arXiv:1611.09855 [hep-ph]}}.

\bibitem{julia}
J.~Bezanson, S.~Karpinski, V.~B. Shah, and A.~Edelman, ``Julia: {A} fast
  dynamic language for technical computing,'' {\em CoRR} {\bfseries
  abs/1209.5145} (2012) , \href{http://arxiv.org/abs/1209.5145}{{\ttfamily
  1209.5145}}. \url{http://arxiv.org/abs/1209.5145}.

\bibitem{StaticArrays}
``Staticarrays.jl.'' \url{https://juliaarrays.github.io/StaticArrays.jl/},
  Aug., 2022.

\bibitem{Pearson1916}
K.~Pearson, ``Mathematical contributions to the theory of evolution. xix.
  second supplement to a memoir on skew variation,'' {\em Philosophical
  Transactions of the Royal Society of London. Series A, Containing Papers of a
  Mathematical or Physical Character} {\bfseries 216} (1916) 429--457.
  \url{http://www.jstor.org/stable/91092}.

\bibitem{SLOAN201695}
J.~Sloan, M.~Hotz, C.~Boutan, {\em et~al.}, ``Limits on axion–photon coupling
  or on local axion density: Dependence on models of the milky way’s dark
  halo,''
  \href{http://dx.doi.org/https://doi.org/10.1016/j.dark.2016.09.003}{{\em
  Physics of the Dark Universe} {\bfseries 14} (2016) 95--102}.
  \url{https://www.sciencedirect.com/science/article/pii/S2212686416300504}.

\bibitem{axionbands2}
V.~Anastassopoulos, F.~Avignone, A.~Bykov, {\em et~al.}, ``Towards a
  medium-scale axion helioscope and haloscope,''
  \href{http://dx.doi.org/10.1088/1748-0221/12/11/P11019}{{\em Journal of
  Instrumentation} {\bfseries 12} (06, 2017) }.

\bibitem{axionbands3}
S.~Scopel, ``Particle dark matter candidates,''
  \href{http://dx.doi.org/10.1088/1742-6596/120/4/042003}{{\em Journal of
  Physics: Conference Series} {\bfseries 120} (07, 2008) 042003}.

\bibitem{Shilon:2012te}
I.~Shilon, A.~Dudarev, H.~Silva, and H.~H.~J. ten Kate, ``{Conceptual Design of
  a New Large Superconducting Toroid for IAXO, the New International AXion
  Observatory},'' \href{http://dx.doi.org/10.1109/TASC.2013.2251052}{{\em IEEE
  Trans. Appl. Supercond.} {\bfseries 23} no.~3, (2013) 4500604},
  \href{http://arxiv.org/abs/1212.4633}{{\ttfamily arXiv:1212.4633
  [physics.ins-det]}}.

\bibitem{CAST:2017uph}
{\bfseries CAST} Collaboration, V.~Anastassopoulos {\em et~al.}, ``{New CAST
  Limit on the Axion-Photon Interaction},''
  \href{http://dx.doi.org/10.1038/nphys4109}{{\em Nature Phys.} {\bfseries 13}
  (2017) 584--590}, \href{http://arxiv.org/abs/1705.02290}{{\ttfamily
  arXiv:1705.02290 [hep-ex]}}.

\bibitem{2019JCAP...06..047A}
E.~{Armengaud}, D.~{Atti{\'e}}, S.~{Basso}, {\em et~al.}, ``{Physics potential
  of the International Axion Observatory (IAXO)},''
  \href{http://dx.doi.org/10.1088/1475-7516/2019/06/047}{{\em JCAP} {\bfseries
  2019} no.~6, (June, 2019) 047},
  \href{http://arxiv.org/abs/1904.09155}{{\ttfamily arXiv:1904.09155
  [hep-ph]}}.

\bibitem{Beurthey:2020yuq}
S.~Beurthey {\em et~al.}, ``{MADMAX Status Report},''
  \href{http://arxiv.org/abs/2003.10894}{{\ttfamily arXiv:2003.10894
  [physics.ins-det]}}.

\bibitem{Grenet:2021vbb}
T.~Grenet, R.~Ballou, Q.~Basto, {\em et~al.}, ``{The Grenoble Axion Haloscope
  platform (GrAHal): development plan and first results},''
  \href{http://arxiv.org/abs/2110.14406}{{\ttfamily arXiv:2110.14406
  [hep-ex]}}.

\bibitem{CAST:2020rlf}
{\bfseries CAST} Collaboration, A.~A. Melc\'on {\em et~al.}, ``{First results
  of the CAST-RADES haloscope search for axions at 34.67 $\mu$eV},''
  \href{http://dx.doi.org/10.1007/JHEP10(2021)075}{{\em JHEP} {\bfseries 21}
  (2020) 075}, \href{http://arxiv.org/abs/2104.13798}{{\ttfamily
  arXiv:2104.13798 [hep-ex]}}.

\bibitem{Crisosto:2019fcj}
N.~Crisosto, P.~Sikivie, N.~S. Sullivan, {\em et~al.}, ``{ADMX SLIC: Results
  from a Superconducting $LC$ Circuit Investigating Cold Axions},''
  \href{http://dx.doi.org/10.1103/PhysRevLett.124.241101}{{\em Phys. Rev.
  Lett.} {\bfseries 124} no.~24, (2020) 241101},
  \href{http://arxiv.org/abs/1911.05772}{{\ttfamily arXiv:1911.05772
  [astro-ph.CO]}}.

\bibitem{Devlin:2021fpq}
J.~A. Devlin {\em et~al.}, ``{Constraints on the Coupling between Axionlike
  Dark Matter and Photons Using an Antiproton Superconducting Tuned Detection
  Circuit in a Cryogenic Penning Trap},''
  \href{http://dx.doi.org/10.1103/PhysRevLett.126.041301}{{\em Phys. Rev.
  Lett.} {\bfseries 126} no.~4, (2021) 041301},
  \href{http://arxiv.org/abs/2101.11290}{{\ttfamily arXiv:2101.11290
  [astro-ph.CO]}}.

\bibitem{QUAX:2020adt}
{\bfseries QUAX} Collaboration, N.~Crescini {\em et~al.}, ``{Axion search with
  a quantum-limited ferromagnetic haloscope},''
  \href{http://dx.doi.org/10.1103/PhysRevLett.124.171801}{{\em Phys. Rev.
  Lett.} {\bfseries 124} no.~17, (2020) 171801},
  \href{http://arxiv.org/abs/2001.08940}{{\ttfamily arXiv:2001.08940
  [hep-ex]}}.

\bibitem{Alesini:2019ajt}
D.~Alesini {\em et~al.}, ``{Galactic axions search with a superconducting
  resonant cavity},'' \href{http://dx.doi.org/10.1103/PhysRevD.99.101101}{{\em
  Phys. Rev. D} {\bfseries 99} no.~10, (2019) 101101},
  \href{http://arxiv.org/abs/1903.06547}{{\ttfamily arXiv:1903.06547
  [physics.ins-det]}}.

\bibitem{Alesini:2020vny}
D.~Alesini {\em et~al.}, ``{Search for invisible axion dark matter of mass
  m$_a=43~\mu$eV with the QUAX--$a\gamma$ experiment},''
  \href{http://dx.doi.org/10.1103/PhysRevD.103.102004}{{\em Phys. Rev. D}
  {\bfseries 103} no.~10, (2021) 102004},
  \href{http://arxiv.org/abs/2012.09498}{{\ttfamily arXiv:2012.09498
  [hep-ex]}}.

\bibitem{Thomson:2019aht}
C.~A. Thomson, B.~T. McAllister, M.~Goryachev, {\em et~al.}, ``{Upconversion
  Loop Oscillator Axion Detection Experiment: A Precision Frequency
  Interferometric Axion Dark Matter Search with a Cylindrical Microwave
  Cavity},'' \href{http://dx.doi.org/10.1103/PhysRevLett.127.019901}{{\em Phys.
  Rev. Lett.} {\bfseries 126} no.~8, (2021) 081803},
  \href{http://arxiv.org/abs/1912.07751}{{\ttfamily arXiv:1912.07751
  [hep-ex]}}. [Erratum: Phys.Rev.Lett. 127, 019901 (2021)].

\bibitem{McAllister:2017lkb}
B.~T. McAllister, G.~Flower, J.~Kruger, {\em et~al.}, ``{The ORGAN Experiment:
  An axion haloscope above 15 GHz},''
  \href{http://dx.doi.org/10.1016/j.dark.2017.09.010}{{\em Phys. Dark Univ.}
  {\bfseries 18} (2017) 67--72},
  \href{http://arxiv.org/abs/1706.00209}{{\ttfamily arXiv:1706.00209
  [physics.ins-det]}}.

\bibitem{Jeong:2020cwz}
J.~Jeong, S.~Youn, S.~Bae, {\em et~al.}, ``{Search for Invisible Axion Dark
  Matter with a Multiple-Cell Haloscope},''
  \href{http://dx.doi.org/10.1103/PhysRevLett.125.221302}{{\em Phys. Rev.
  Lett.} {\bfseries 125} no.~22, (2020) 221302},
  \href{http://arxiv.org/abs/2008.10141}{{\ttfamily arXiv:2008.10141
  [hep-ex]}}.

\bibitem{Gramolin:2020ict}
A.~V. Gramolin, D.~Aybas, D.~Johnson, {\em et~al.}, ``{Search for axion-like
  dark matter with ferromagnets},''
  \href{http://dx.doi.org/10.1038/s41567-020-1006-6}{{\em Nature Phys.}
  {\bfseries 17} no.~1, (2021) 79--84},
  \href{http://arxiv.org/abs/2003.03348}{{\ttfamily arXiv:2003.03348
  [hep-ex]}}.

\bibitem{Salemi:2021gck}
C.~P. Salemi {\em et~al.}, ``{Search for Low-Mass Axion Dark Matter with
  ABRACADABRA-10~cm},''
  \href{http://dx.doi.org/10.1103/PhysRevLett.127.081801}{{\em Phys. Rev.
  Lett.} {\bfseries 127} no.~8, (2021) 081801},
  \href{http://arxiv.org/abs/2102.06722}{{\ttfamily arXiv:2102.06722
  [hep-ex]}}.

\bibitem{Ouellet:2018beu}
J.~L. Ouellet {\em et~al.}, ``{First Results from ABRACADABRA-10 cm: A Search
  for Sub-$\mu$eV Axion Dark Matter},''
  \href{http://dx.doi.org/10.1103/PhysRevLett.122.121802}{{\em Phys. Rev.
  Lett.} {\bfseries 122} no.~12, (2019) 121802},
  \href{http://arxiv.org/abs/1810.12257}{{\ttfamily arXiv:1810.12257
  [hep-ex]}}.

\bibitem{HAYSTAC:2018rwy}
{\bfseries HAYSTAC} Collaboration, L.~Zhong {\em et~al.}, ``{Results from phase
  1 of the HAYSTAC microwave cavity axion experiment},''
  \href{http://dx.doi.org/10.1103/PhysRevD.97.092001}{{\em Phys. Rev. D}
  {\bfseries 97} no.~9, (2018) 092001},
  \href{http://arxiv.org/abs/1803.03690}{{\ttfamily arXiv:1803.03690
  [hep-ex]}}.

\bibitem{HAYSTAC:2020kwv}
{\bfseries HAYSTAC} Collaboration, K.~M. Backes {\em et~al.}, ``{A
  quantum-enhanced search for dark matter axions},''
  \href{http://dx.doi.org/10.1038/s41586-021-03226-7}{{\em Nature} {\bfseries
  590} no.~7845, (2021) 238--242},
  \href{http://arxiv.org/abs/2008.01853}{{\ttfamily arXiv:2008.01853
  [quant-ph]}}.

\bibitem{hagmann1990results}
C.~Hagmann, P.~Sikivie, N.~Sullivan, and D.~Tanner, ``Results from a search for
  cosmic axions,'' {\em Physical Review D} {\bfseries 42} no.~4, (1990) 1297.

\bibitem{depanfilis1987limits}
S.~DePanfilis, A.~Melissinos, B.~Moskowitz, {\em et~al.}, ``Limits on the
  abundance and coupling of cosmic axions at $4.5 < m_a < 5.0 \; \mu$ev,'' {\em
  Physical Review Letters} {\bfseries 59} no.~7, (1987) 839.

\bibitem{ADMX:2018gho}
{\bfseries ADMX} Collaboration, N.~Du {\em et~al.}, ``{A Search for Invisible
  Axion Dark Matter with the Axion Dark Matter Experiment},''
  \href{http://dx.doi.org/10.1103/PhysRevLett.120.151301}{{\em Phys. Rev.
  Lett.} {\bfseries 120} no.~15, (2018) 151301},
  \href{http://arxiv.org/abs/1804.05750}{{\ttfamily arXiv:1804.05750
  [hep-ex]}}.

\bibitem{ADMX:2019uok}
{\bfseries ADMX} Collaboration, T.~Braine {\em et~al.}, ``{Extended Search for
  the Invisible Axion with the Axion Dark Matter Experiment},''
  \href{http://dx.doi.org/10.1103/PhysRevLett.124.101303}{{\em Phys. Rev.
  Lett.} {\bfseries 124} no.~10, (2020) 101303},
  \href{http://arxiv.org/abs/1910.08638}{{\ttfamily arXiv:1910.08638
  [hep-ex]}}.

\bibitem{ADMX:2021nhd}
{\bfseries ADMX} Collaboration, C.~Bartram {\em et~al.}, ``{Search for
  Invisible Axion Dark Matter in the 3.3\textendash{}4.2\,\,\ensuremath{\mu}eV
  Mass Range},'' \href{http://dx.doi.org/10.1103/PhysRevLett.127.261803}{{\em
  Phys. Rev. Lett.} {\bfseries 127} no.~26, (2021) 261803},
  \href{http://arxiv.org/abs/2110.06096}{{\ttfamily arXiv:2110.06096
  [hep-ex]}}.

\bibitem{Baryakhtar:2018doz}
M.~Baryakhtar, J.~Huang, and R.~Lasenby, ``{Axion and hidden photon dark matter
  detection with multilayer optical haloscopes},''
  \href{http://dx.doi.org/10.1103/PhysRevD.98.035006}{{\em Phys. Rev. D}
  {\bfseries 98} no.~3, (2018) 035006},
  \href{http://arxiv.org/abs/1803.11455}{{\ttfamily arXiv:1803.11455
  [hep-ph]}}.

\bibitem{Michimura:2019qxr}
Y.~Michimura, Y.~Oshima, T.~Watanabe, {\em et~al.}, ``{DANCE: Dark matter Axion
  search with riNg Cavity Experiment},''
  \href{http://dx.doi.org/10.1088/1742-6596/1468/1/012032}{{\em J. Phys. Conf.
  Ser.} {\bfseries 1468} no.~1, (2020) 012032},
  \href{http://arxiv.org/abs/1911.05196}{{\ttfamily arXiv:1911.05196
  [physics.ins-det]}}.

\bibitem{Aja:2022csb}
B.~Aja {\em et~al.}, ``{The Canfranc Axion Detection Experiment (CADEx): search
  for axions at 90 GHz with Kinetic Inductance Detectors},''
  \href{http://dx.doi.org/10.1088/1475-7516/2022/11/044}{{\em JCAP} {\bfseries
  11} (2022) 044}, \href{http://arxiv.org/abs/2206.02980}{{\ttfamily
  arXiv:2206.02980 [hep-ex]}}.

\bibitem{BREAD:2021tpx}
{\bfseries BREAD} Collaboration, J.~Liu {\em et~al.}, ``{Broadband Solenoidal
  Haloscope for Terahertz Axion Detection},''
  \href{http://dx.doi.org/10.1103/PhysRevLett.128.131801}{{\em Phys. Rev.
  Lett.} {\bfseries 128} no.~13, (2022) 131801},
  \href{http://arxiv.org/abs/2111.12103}{{\ttfamily arXiv:2111.12103
  [physics.ins-det]}}.

\bibitem{Millar:2022peq}
A.~J. Millar {\em et~al.}, ``{ALPHA: Searching For Dark Matter with Plasma
  Haloscopes},'' \href{http://arxiv.org/abs/2210.00017}{{\ttfamily
  arXiv:2210.00017 [hep-ph]}}.

\bibitem{Lawson:2019brd}
M.~Lawson, A.~J. Millar, M.~Pancaldi, {\em et~al.}, ``{Tunable axion plasma
  haloscopes},'' \href{http://dx.doi.org/10.1103/PhysRevLett.123.141802}{{\em
  Phys. Rev. Lett.} {\bfseries 123} no.~14, (2019) 141802},
  \href{http://arxiv.org/abs/1904.11872}{{\ttfamily arXiv:1904.11872
  [hep-ph]}}.

\bibitem{TASEH:2022vvu}
{\bfseries TASEH} Collaboration, H.~Chang {\em et~al.}, ``{First Results from
  the Taiwan Axion Search Experiment with a Haloscope at
  19.6\,\,\ensuremath{\mu}eV},''
  \href{http://dx.doi.org/10.1103/PhysRevLett.129.111802}{{\em Phys. Rev.
  Lett.} {\bfseries 129} no.~11, (2022) 111802},
  \href{http://arxiv.org/abs/2205.05574}{{\ttfamily arXiv:2205.05574
  [hep-ex]}}.

\bibitem{2022PhRvD.106e2007A}
D.~{Alesini}, D.~{Babusci}, C.~{Braggio}, {\em et~al.}, ``{Search for Galactic
  axions with a high-Q dielectric cavity},''
  \href{http://dx.doi.org/10.1103/PhysRevD.106.052007}{{\em Phys. Rev. D}
  {\bfseries 106} no.~5, (Sept., 2022) 052007},
  \href{http://arxiv.org/abs/2208.12670}{{\ttfamily arXiv:2208.12670
  [hep-ex]}}.

\bibitem{2022arXiv220312152Q}
A.~P. {Quiskamp}, B.~T. {McAllister}, P.~{Altin}, {\em et~al.}, ``{Direct
  Search for Dark Matter Axions Excluding ALP Cogenesis in the 63-67 micro-eV
  Range, with The ORGAN Experiment},''
  \href{http://dx.doi.org/10.48550/arXiv.2203.12152}{{\em arXiv e-prints}
  (Mar., 2022) arXiv:2203.12152},
  \href{http://arxiv.org/abs/2203.12152}{{\ttfamily arXiv:2203.12152
  [hep-ex]}}.

\bibitem{2023arXiv230109721H}
{HAYSTAC Collaboration}, M.~J. {Jewell}, A.~F. {Leder}, {\em et~al.}, ``{New
  Results from HAYSTAC's Phase II Operation with a Squeezed State Receiver},''
  \href{http://dx.doi.org/10.48550/arXiv.2301.09721}{{\em arXiv e-prints}
  (Jan., 2023) arXiv:2301.09721},
  \href{http://arxiv.org/abs/2301.09721}{{\ttfamily arXiv:2301.09721
  [hep-ex]}}.

\bibitem{2020PhRvL.124j1802L}
S.~{Lee}, S.~{Ahn}, J.~{Choi}, {\em et~al.}, ``{Axion Dark Matter Search around
  6.7 {\ensuremath{\mu}} eV},''
  \href{http://dx.doi.org/10.1103/PhysRevLett.124.101802}{{\em Phys. Rev.
  Lett.} {\bfseries 124} no.~10, (Mar., 2020) 101802},
  \href{http://arxiv.org/abs/2001.05102}{{\ttfamily arXiv:2001.05102
  [hep-ex]}}.

\bibitem{2021PhRvL.126s1802K}
O.~{Kwon}, D.~{Lee}, W.~{Chung}, {\em et~al.}, ``{First Results from an Axion
  Haloscope at CAPP around 10.7 {\ensuremath{\mu}} eV},''
  \href{http://dx.doi.org/10.1103/PhysRevLett.126.191802}{{\em Phys. Rev.
  Lett.} {\bfseries 126} no.~19, (May, 2021) 191802},
  \href{http://arxiv.org/abs/2012.10764}{{\ttfamily arXiv:2012.10764
  [hep-ex]}}.

\bibitem{2022PhRvL.128x1805L}
Y.~{Lee}, B.~{Yang}, H.~{Yoon}, {\em et~al.}, ``{Searching for Invisible Axion
  Dark Matter with an 18 T Magnet Haloscope},''
  \href{http://dx.doi.org/10.1103/PhysRevLett.128.241805}{{\em Phys. Rev.
  Lett.} {\bfseries 128} no.~24, (June, 2022) 241805},
  \href{http://arxiv.org/abs/2206.08845}{{\ttfamily arXiv:2206.08845
  [hep-ex]}}.

\bibitem{2022arXiv220713597K}
J.~{Kim}, O.~{Kwon}, {\c{C}}.~{Kutlu}, {\em et~al.}, ``{Near-Quantum-Noise
  Axion Dark Matter Search at CAPP around 9.5 $\mu$eV},''
  \href{http://dx.doi.org/10.48550/arXiv.2207.13597}{{\em arXiv e-prints}
  (July, 2022) arXiv:2207.13597},
  \href{http://arxiv.org/abs/2207.13597}{{\ttfamily arXiv:2207.13597
  [hep-ex]}}.

\bibitem{2022arXiv221010961Y}
A.~K. {Yi}, S.~{Ahn}, {\c{C}}.~{Kutlu}, {\em et~al.}, ``{Axion Dark Matter
  Search around 4.55 $\mu$eV with Dine-Fischler-Srednicki-Zhitnitskii
  Sensitivity},'' \href{http://dx.doi.org/10.48550/arXiv.2210.10961}{{\em arXiv
  e-prints} (Oct., 2022) arXiv:2210.10961},
  \href{http://arxiv.org/abs/2210.10961}{{\ttfamily arXiv:2210.10961
  [hep-ex]}}.

\bibitem{Berlin:2020vrk}
A.~Berlin, R.~T. D'Agnolo, S.~A.~R. Ellis, and K.~Zhou, ``{Heterodyne broadband
  detection of axion dark matter},''
  \href{http://dx.doi.org/10.1103/PhysRevD.104.L111701}{{\em Phys. Rev. D}
  {\bfseries 104} no.~11, (2021) L111701},
  \href{http://arxiv.org/abs/2007.15656}{{\ttfamily arXiv:2007.15656
  [hep-ph]}}.

\bibitem{DMRadio:2022pkf}
{\bfseries DMRadio} Collaboration, L.~Brouwer {\em et~al.}, ``{Projected
  sensitivity of DMRadio-m3: A search for the QCD axion below
  1\,\,\ensuremath{\mu}eV},''
  \href{http://dx.doi.org/10.1103/PhysRevD.106.103008}{{\em Phys. Rev. D}
  {\bfseries 106} no.~10, (2022) 103008},
  \href{http://arxiv.org/abs/2204.13781}{{\ttfamily arXiv:2204.13781
  [hep-ex]}}.

\bibitem{Alesini:2017ifp}
D.~Alesini, D.~Babusci, D.~Di~Gioacchino, {\em et~al.}, ``{The KLASH
  Proposal},'' \href{http://arxiv.org/abs/1707.06010}{{\ttfamily
  arXiv:1707.06010 [physics.ins-det]}}.

\bibitem{Zhang:2021bpa}
Z.~Zhang, D.~Horns, and O.~Ghosh, ``{Search for dark matter with an LC
  circuit},'' \href{http://dx.doi.org/10.1103/PhysRevD.106.023003}{{\em Phys.
  Rev. D} {\bfseries 106} no.~2, (2022) 023003},
  \href{http://arxiv.org/abs/2111.04541}{{\ttfamily arXiv:2111.04541
  [hep-ex]}}.

\bibitem{Nagano:2019rbw}
K.~Nagano, T.~Fujita, Y.~Michimura, and I.~Obata, ``{Axion Dark Matter Search
  with Interferometric Gravitational Wave Detectors},''
  \href{http://dx.doi.org/10.1103/PhysRevLett.123.111301}{{\em Phys. Rev.
  Lett.} {\bfseries 123} no.~11, (2019) 111301},
  \href{http://arxiv.org/abs/1903.02017}{{\ttfamily arXiv:1903.02017
  [hep-ph]}}.

\bibitem{Liu:2018icu}
H.~Liu, B.~D. Elwood, M.~Evans, and J.~Thaler, ``{Searching for Axion Dark
  Matter with Birefringent Cavities},''
  \href{http://dx.doi.org/10.1103/PhysRevD.100.023548}{{\em Phys. Rev. D}
  {\bfseries 100} no.~2, (2019) 023548},
  \href{http://arxiv.org/abs/1809.01656}{{\ttfamily arXiv:1809.01656
  [hep-ph]}}.

\bibitem{Schutte-Engel:2021bqm}
J.~Sch\"utte-Engel, D.~J.~E. Marsh, A.~J. Millar, {\em et~al.}, ``{Axion
  quasiparticles for axion dark matter detection},''
  \href{http://dx.doi.org/10.1088/1475-7516/2021/08/066}{{\em JCAP} {\bfseries
  08} (2021) 066}, \href{http://arxiv.org/abs/2102.05366}{{\ttfamily
  arXiv:2102.05366 [hep-ph]}}.

\bibitem{Marsh:2018dlj}
D.~J.~E. Marsh, K.-C. Fong, E.~W. Lentz, {\em et~al.}, ``{Proposal to Detect
  Dark Matter using Axionic Topological Antiferromagnets},''
  \href{http://dx.doi.org/10.1103/PhysRevLett.123.121601}{{\em Phys. Rev.
  Lett.} {\bfseries 123} no.~12, (2019) 121601},
  \href{http://arxiv.org/abs/1807.08810}{{\ttfamily arXiv:1807.08810
  [hep-ph]}}.

\bibitem{AxionLimits}
C.~O'Hare, ``cajohare/axionlimits: Axionlimits.''
  \url{https://cajohare.github.io/AxionLimits/}, Jan., 2023.

\bibitem{Bjorkeroth:2019jtx}
F.~Bj\"orkeroth, L.~Di~Luzio, F.~Mescia, {\em et~al.}, ``{Axion-electron
  decoupling in nucleophobic axion models},''
  \href{http://dx.doi.org/10.1103/PhysRevD.101.035027}{{\em Phys. Rev. D}
  {\bfseries 101} no.~3, (2020) 035027},
  \href{http://arxiv.org/abs/1907.06575}{{\ttfamily arXiv:1907.06575
  [hep-ph]}}.

\bibitem{Dvali:2005an}
G.~Dvali, ``{Three-form gauging of axion symmetries and gravity},''
  \href{http://arxiv.org/abs/hep-th/0507215}{{\ttfamily arXiv:hep-th/0507215}}.

\bibitem{PhysRevD.105.085020}
O.~Sakhelashvili, ``Consistency of the dual formulation of axion solutions to
  the strong $cp$ problem,''
  \href{http://dx.doi.org/10.1103/PhysRevD.105.085020}{{\em Phys. Rev. D}
  {\bfseries 105} (Apr, 2022) 085020}.
  \url{https://link.aps.org/doi/10.1103/PhysRevD.105.085020}.

\end{thebibliography}\endgroup

\end{document}